\documentclass[usenatbib]{mnras}

\newcommand\lsim{\mathrel{\rlap{\lower4pt\hbox{\hskip1pt$\sim$}}
        \raise1pt\hbox{$<$}}}
\newcommand\gsim{\mathrel{\rlap{\lower4pt\hbox{\hskip1pt$\sim$}}
        \raise1pt\hbox{$>$}}}

\usepackage[usenames, dvipsnames]{color}
\usepackage{graphicx}
\usepackage{verbatim}

\usepackage{hyperref}
\hypersetup{
    pdfnewwindow=true,      
    colorlinks=true,       
    linkcolor=blue,          
    citecolor=blue,        
    filecolor=blue,      
    urlcolor=blue           
}

\usepackage{calligra}
\DeclareMathAlphabet{\mathcalligra}{T1}{calligra}{m}{n}
\DeclareFontShape{T1}{calligra}{m}{n}{<->s*[2.2]callig15}{}

\def\prd{PRD}
\def\apj{ApJ}                 
\def\apjl{ApJL}               
\def\apjs{ApJS}               
\def\mnras{MNRAS}             
\def\aap{A\&A}                
\def\nat{Nature}              
\def\araa{ARA\&A}             
\def\actaa{Acta Astron}  
\def\icarus{Icarus}  

\def\Mach{\mathcal{M}}
\def\bin{\rm{bin}}

\begin{document}

\title[A transition in circumbinary discs]{A transition in circumbinary accretion discs at a binary mass ratio of 1:25}

\author[D. J. D'Orazio, Z. Haiman, P. Duffell, A. I. MacFadyen, B. D. Farris]{Daniel J. D'Orazio$^1$,
  Zolt\'an~Haiman$^1$,
  Paul Duffell$^2$, 
  Andrew MacFadyen$^3$, 
  \and Brian Farris$^{1,3}$  \\
  \thanks{dorazio@astro.columbia.edu; zoltan@astro.columbia.edu}\\
     $^1$Department of Astronomy, Columbia University, 550 West 120th Street, New York, NY 10027 \\
     $^2$Theoretical Astrophysics Center, University of California, Berkeley \\
     $^3$Department of Physics, New York University}

\maketitle

\begin{abstract}
We study circumbinary accretion discs in the framework of the
restricted three-body problem (R3Bp) and via numerically solving the
height-integrated equations of viscous hydrodynamics. Varying the mass
ratio of the binary, we find a pronounced change in the behaviour of
the disc near mass ratio $q \equiv M_s/M_p \sim 0.04$.  For mass
ratios above $q=0.04$, solutions for the hydrodynamic flow transition
from steady, to strongly-fluctuating; a narrow annular gap in the
surface density around the secondary's orbit changes to a hollow
central cavity; and a spatial symmetry is lost, resulting in a
lopsided disc. This phase transition is coincident with the mass ratio
above which stable orbits do not exist around the L4 and L5
equilibrium points of the R3B problem.  Using the DISCO code, we find
that for thin discs, for which a gap or cavity can remain open, the
mass ratio of the transition is relatively insensitive to disc
viscosity and pressure.  The $q=0.04$ transition has relevance for the
evolution of massive black hole binary+disc systems at the centers of
galactic nuclei, as well as for young stellar binaries and possibly
planets around brown dwarfs.
\end{abstract}

\section{Introduction}
Binaries embedded in gas discs are ubiquitous astrophysical systems. They are realized in the proto-planetary nebulae surrounding young stars and their growing planets \citep{KleyNelson:2012:rev} and possibly in young binary star systems as evidenced by circumbinary planets \citep[\textit{e.g.}][]{ Orosz:2012Sci}. They also arise at the centers of galactic nuclei to which gas can be funneled to accompany an inspiraling massive black hole binary (MBHB) \citep[][and see recent reviews by \cite{Dotti:2012:rev,Mayer:2013:MBHBGasRev}]{Barnes:1996}.

Understanding the long-term evolution of the binary+disc system is
complicated by the coupled nature of mass, angular momentum, and energy
conservation for the total binary+disc system. The binary affects the
structure of the disc, and the disc alters the orbital parameters of
the binary. For planets and stars enveloped by a gas disc, the
binary+disc interaction determines the migration and growth of the
planets, dictating the post-disc-configuration of the planetary
system. For a MBHB+disc system, gas torques can alter the inspiral
rate of the binary. The effect is important for deciphering the final
parsec problem and predicting the rate of gravitational wave events
due to MBHB mergers \citep{Begel:Blan:Rees:1980, GouldRix:2000,
  ArmNat:2002:ApJL, ArmNat:2005}, and possibly even affecting the
gravitational wave signal from inspiral \citep[D'Orazio et al., {\em in preparation};][]{YKH:2011:L, KocsisYunesLoeb:2011}.

Additionally, interaction of the binary and disc can lead to periodic accretion \citep{Hayasaki:2007, MacFadyen:2008, Cuadra:2009, Roedig:2011:eccevo, Noble+2012,ShiKrolik:2012:ApJ, Roedig:2012:Trqs, DHM:2013:MNRAS, Farris:2014, Dunhill:2015, ShiKrolik:2015} which can aid in identifying MBHB candidates in electromagnetic (EM) surveys \citep{HKM09}. As has been recently been made clear by the discovery of multiple MBHB candidates in EM time-domain surveys \citep{Graham+2015a, Graham+2015b, Liu:7RsMBHB:2015}, the interpretation of variability in EM surveys will rely heavily on our knowledge of how accretion variability depends on system parameters such as the binary mass ratio \citep[see \textit{e.g.}][]{PG1302MNRAS:2015a,PG1302Nature:2015b}.

Although the disc and binary are coupled, a useful first step in
determining their mutual evolution is to determine the perturbation to
the disc surface density by a binary on a fixed orbit. From this
exercise three distinct regimes arise as a function of binary mass 
ratio ($q \equiv M_s/M_p$) and disc hydrodynamic parameters. 
Small-mass-ratio binaries, $q_{\rm{lin}} \lsim \Mach^{-5/2}\alpha^{1/2}$
\citep{DuffellMac:2013:smallqGapOpen} excite linear spiral density
waves in the disc. Here $\Mach$ is the disc Mach number near the
binary's orbit and $\alpha$ is the alpha-law viscosity parameter
\citep{SS73}. The torque on the binary from the spiral density wave
perturbation causes the binary's orbit to shrink on the so-called Type
I migration timescale \citep{GT79, GT80, Ward:1997}.

Larger-mass-ratio binaries ($q \gsim \Mach^{-5/2}\alpha^{1/2}$) open a low surface density, annular gap in the disc, altering the binary migration rate. Often it is assumed that the migration in this regime is equal to the viscous drift rate in the disc \citep{LinPapa86b, NelsonKley:2000, DAngelowLubow:2008}, though recent numerical works have called this into question  \citep{Edgar:2008, DuffellFTV:2014, DurmannKley:2015}.

The critical mass ratio for gap opening depends on the Mach number and disc viscosity. In the small mass ratio regime, this criterion has been explored analytically \citep[{\em e.g.},][]{GT79, GT80, PapaPringle:1977} and numerically \citep[{\em e.g.},][]{Bryden:1999, NelsonKley:2000,Papa:2004:MHDGapopen, Zhu:2013:GapOpen}. It has been thought that a necessary condition for a gap is that the secondary's Hill radius be larger than the disc scale height \citep{LinPapa:1993:ConfProcRev, CMM:2006}. \cite{GoodmanRafikov:2001} first proposed that this was not a necessary condition, but that low-mass perturbers could open gaps in low-viscosity discs, on much longer timescales. This has so far been validated in 2D simulations \citep{DongRafII:2011, DM2012:gaps, DuffellMac:2013:smallqGapOpen, FungGaps:2014}, though it has not yet been validated in 3D, due to the computational expense and long timescales necessary to capture gap opening in the low-mass-ratio regime.

For binary mass ratios near unity, hydrodynamical simulations find a 
central, low-density cavity cleared in the disc 
\citep[{\em e.g.},][]{Artymowicz:1994, ArtyLubow:1996, Farris:2014}. 
A variety of methods have been employed to determine the properties 
of circumbinary discs (CBDs) with near-equal binary masses. 
Cavity sizes have been estimated by calculating the truncation radius of the 
inner edge of CBDs, determined by orbit intersection 
and stability in the restricted three-body problem (R3Bp) \citep{RP:Excretion:1981} 
and through resonant torque calculations \citep{Artymowicz:1994, 
ArtyLubow:1996}. Studies by \cite{delValleEscala:I:2012} and \cite{delValleEscala:II:2014} have examined 
cavity opening/closing conditions for $q>0.1$ binaries with massive discs by 
calculating non-resonant torques due to a non-axisymmetric disc structure. 
\cite{Roedig:2012:Trqs} have used 3D, 
smoothed-particle hydrodynamics to analyze the gas and binary
torques acting on a massive disc and a near-unity, $q=1/3$, 
mass ratio binary. They find binary orbital decay and binary eccentricity 
growth in the presence of a central cavity fed by gas streams.
None of these studies, however, has asked what conditions are 
required to form a central cavity rather than an annular gap, 
or even if an important distinction exists between the two regimes.

While the transition from the small mass ratio, weakly-perturbed
regime to the larger mass ratio, gapped regime is well defined in the
literature, the transition to the near-unity mass ratio state is
not. Here we utilize the circular R3Bp as well as 2D viscous
hydrodynamical simulations to show that there is a dynamically
important transition from the gapped regime to the near-unity mass
ratio regime which is marked by
\begin{enumerate}
\item A transition in surface density structure from a low-density
  annular gap towards a lower-density central cavity.
\item A transition from steady-state to strongly-fluctuating disc
  dynamics.
\item The development of strong asymmetry (i.e. a lopsided shape) of
  the central cavity and the slow precession of this cavity.
\end{enumerate}
These characteristics of large binary mass ratio systems begin to
appear above a mass ratio of $q \sim 0.04$. This is the same mass
ratio above which stable orbits cease to exist in the co-rotation region
of the R3Bp \citep{MD:SSD}. Here we show that the above transition 
occurs over a wide range of hydrodynamical parameters and provide 
evidence that the transition is linked to R3B-orbital-stability criteria.

This study proceeds as follows. In \S \ref{Stability Analysis} we use
the integral of motion of the R3Bp to infer the structure of density
gaps and cavities in a CBD. In \S \ref{Integration
  of the equations of motion} we integrate the equations of the
circular R3Bp to elaborate on these findings. \S \ref{Viscous and
  Pressure Effects} analytically considers the effects of pressure and
viscosity which are present for an astrophysical CBD. In \S
\ref{Fiducial Simulations}, we present viscous hydrodynamical
simulations to compare with the findings of the R3Bp analysis. In \S
\ref{ParamStudy} we conduct a parameter study over disc viscosity and
pressure, providing evidence that the high mass ratio, CBD transition
is generated by the loss of stable particle orbits in the binary
co-orbital region.

\section{Restricted 3-body Analysis}
\label{Analysis via the restricted 3-body problem}
Gap/cavity clearing and morphology are governed by the gravitational
interaction between disc particles and the binary, as well as viscous
and pressure forces. We conduct a purely gravitational study by
ignoring hydrodynamical effects, treating the disc as a collection of
test-particles obeying the circular R3Bp equations of motion (See
\textit{e.g.} \cite{MD:SSD}). This gravitational study lends
remarkable insight into the full hydrodynamical problem.

\subsection{Restrictions on Orbits from Conserved Integrals of Motion}
\label{Stability Analysis}
\begin{figure}
\begin{center}
\includegraphics[scale=0.55]{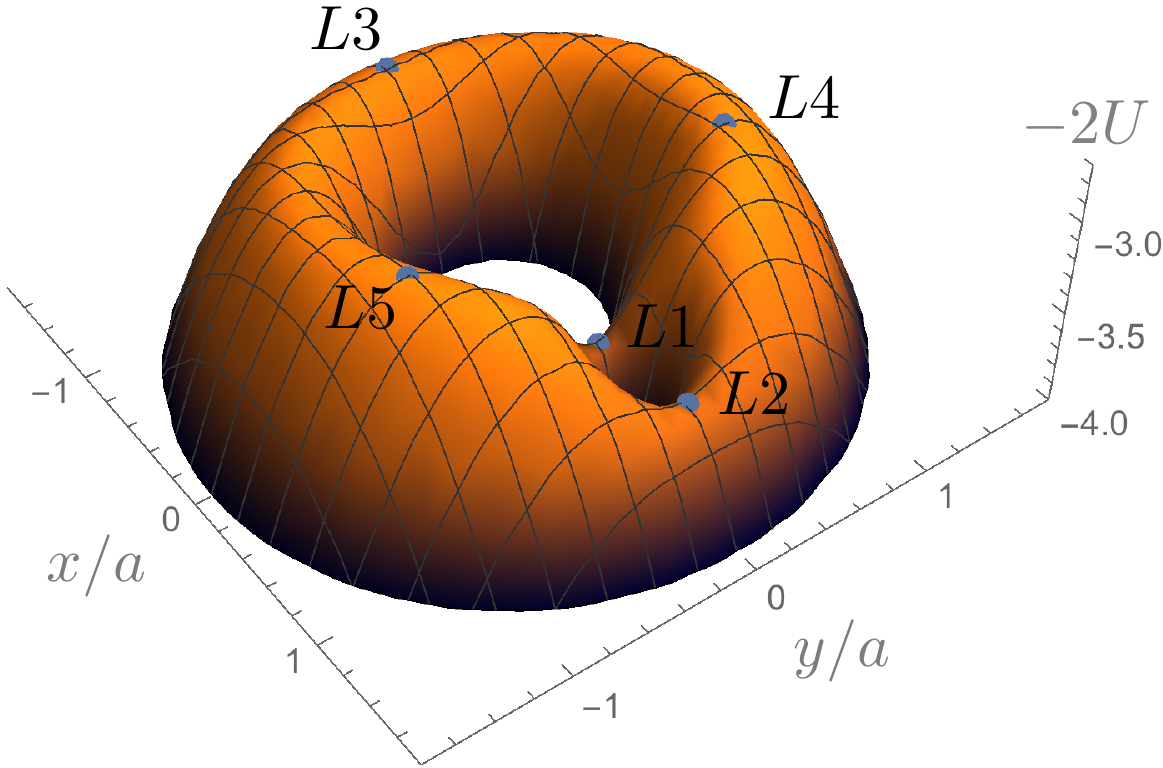} 
\end{center}
\caption{A three-dimensional representation of the effective binary
  potential in the co-rotating frame for a binary with mass ratio
  $q=0.1$. Here we have plotted twice the Roche potential, $-2U$, the negative of the Jacobi
  constant for a particle with zero velocity (see Eq. \ref{Eq:CJ}). The
  x and y coordinates in the binary plane are measured in units of the
  binary separation $a$. The primary and secondary are located at
  $(x_p,y_p) = (-a/(1+1/q), 0)$ and $(x_s,y_s) = (a/(1+q), 0)$
  respectively. The five Lagrange points are labeled for reference.}
\label{Fig:Roche3D}
\end{figure}

Circumbinary discs are characterized by global, low-density 
gaps and cavities. These structures are created
when gas/particles cannot stably exist in a region. We begin by
searching for restricted regions in the binary-disc plane. We seek to
find which particles in the binary orbital plane are restricted from
crossing into/out of the binary's orbit, and which are free to be
expelled in the formation of a gap or cavity. To do this we utilize
the Jacobi constant
\begin{equation}
C_J = 2U - v^2, 
\label{Eq:CJ}
\end{equation}
where $U$ is the negative of the Roche potential of the binary depicted in
Figure \ref{Fig:Roche3D}, $v$ is the velocity of a test particle orbiting the binary, and
all quantities are functions of the coordinates. As the only integral
of motion in the R3Bp, the Jacobi constant is conserved along a
test-particle orbit. Since $C_J$ is conserved, a particle with Jacobi
constant $C^p_J$ is restricted to regions of the binary orbital plane
where $C^p_J \leq 2U$, else the particle would have a complex
velocity. We use this property to draw zero-velocity curves (ZVCs),
level curves of the Roche potential (see Figure \ref{Fig:Roche3D}), defined
by the equation $C^*_J = 2U$. Particles with $C_J > C^*_J$ cannot
enter the closed region delineated by the ZVCs.

\begin{figure}
\begin{center}
\includegraphics[scale=0.55]{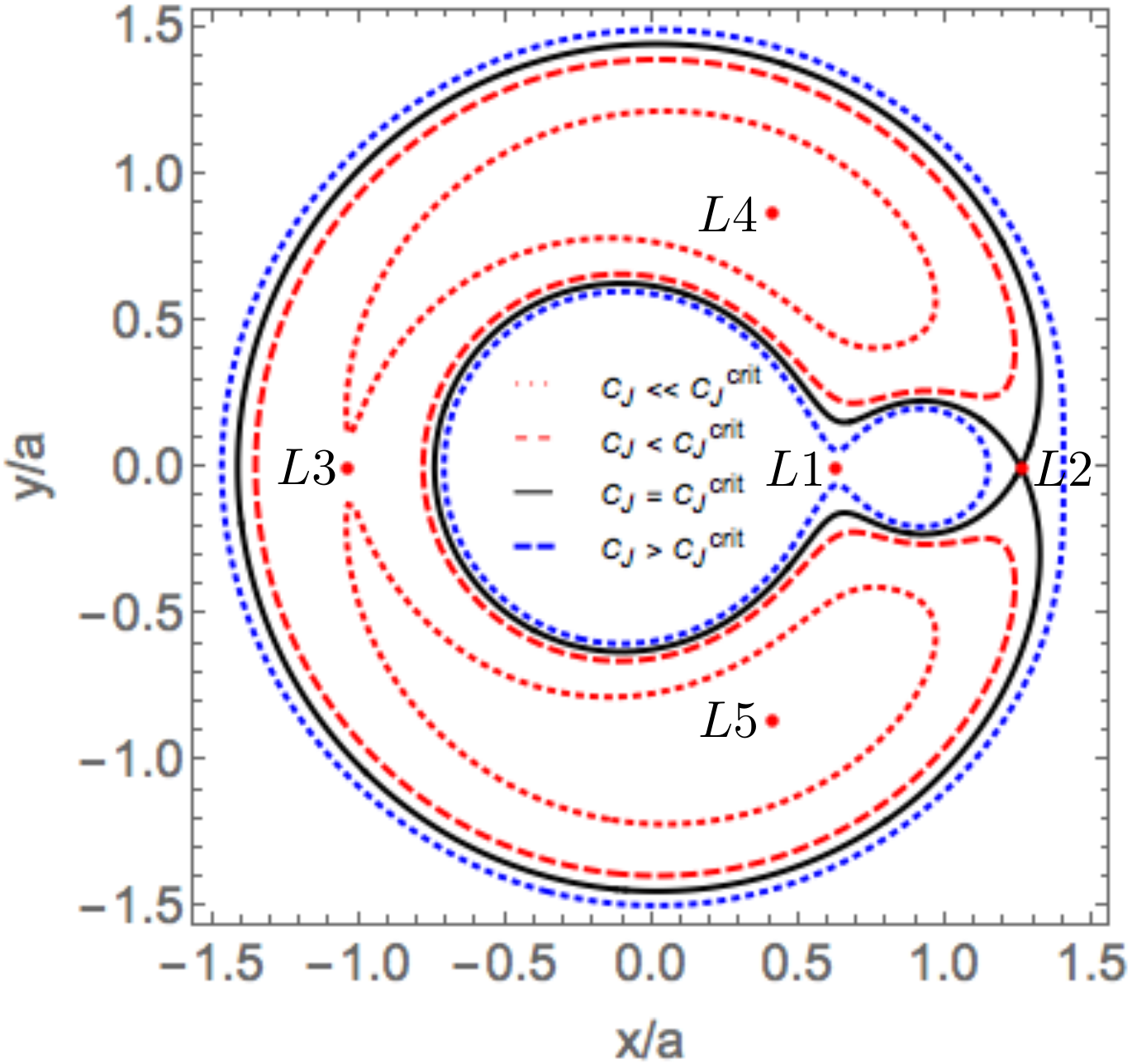} 
\end{center}
\caption{Zero-velocity curves for four different values of the Jacobi
  constant for a binary with mass ratio $q=M_s/M_p=0.1$ (level curves
  of the potential plotted in Figure \ref{Fig:Roche3D}). For $C_J
  \geq C^{\rm{crit}}$ (blue thick dashed and solid black), the
  zero-velocity curves connect, separating the binary plane into
  distinct inner and outer regions (as well as a third region around
  the secondary for large enough values of the Jacobi Constant). For
  $C_J < C^{\rm{crit}}$ (red lines), the zero-velocity curves open at
  L2 and then at L3 for even smaller $C_J$. The critical zero-velocity
  curve (black) passes through the Lagrange point L2. The five
  Lagrange points are labeled for reference.}
\label{Fig:CJ_Ex}
\end{figure}

For the purpose of studying gap/cavity structure, we examine the ZVCs
which connect, \textit{i.e} separate the binary plane into $\geq 2$
distinct regions, an inner disc and an outer disc. We define a
critical ZVC corresponding to the smallest value of the Jacobi
constant for which ZVCs connect, $C^{\rm{crit}}_J$. This is the Jacobi
constant with ZVC which passes through the second Lagrange point,
L2. Particles in the disc for which $C_J > C^{\rm{crit}}_J$ cannot
pass between the inner and outer disc.  ZVCs for values of $C_J$ at
and around the critical value are illustrated in Figure
\ref{Fig:CJ_Ex} for a binary with mass ratio $q\equiv M_s/M_p
=0.1$.

To determine the structure of the R3B disc, we ask if particles in the
inner disc have $C_J < C^{\rm{crit}}_J$; if so, they may be evacuated
by the binary to form a central cavity. If the majority of particles
in the inner (outer) disc have $C_J > C^{\rm{crit}}_J$, then they are
trapped in the inner (outer) disc, and an annular gap will define the
density structure. This analysis does not determine whether the un-trapped orbits
actually cross the ZVCs or not - this depends on the initial magnitude
and orientation of the velocity vector of each particle that is not trapped by the 
critical ZVC (green regions of Figure \ref{Fig:CJAddV} below). We determine the 
fate of the un-trapped particles in \S\ref{Integration of the equations of motion}.

Since the value of $C_J$ depends on a particle's position as well as
its velocity, we must prescribe a velocity profile in the disc. We
choose a prescription given by the virial theorem which approaches the
Keplerian value for the binary at large $r\equiv\sqrt{x^2 + y^2}$ and
the Keplerian value for each BH at $r \rightarrow r_p$ and $r
\rightarrow r_s$
\begin{equation}
v_{\phi} =   \sqrt{\frac{GM_s}{r_s} + \frac{GM_p}{r_p}}  - r\Omega_{\bin}.
\label{Eq:AddV}
\end{equation}
where $r_p$ and $r_s$ are the $\phi$ dependent distances from the
primary and secondary, and we subtract the angular frequency
$\Omega_{\bin}$ because we are working in the rotating frame of the
binary.

Given the velocity profile in Eq. (\ref{Eq:AddV}), Figure
\ref{Fig:CJAddV} displays the morphology of restricted regions in the
binary orbital plane. Particles trapped in the outer (inner) disc have
Jacobi constant greater than the critical value and are painted blue
(red). Disc particles which are not restricted to either region are
painted green. The enclosed area of the critical ZVC, which also
consists of un-trapped particles, is painted dark-green.

We emphasize that the shape of the restricted regions depends on the
choice of velocity profile, while the ZVCs are independent of this
choice. An extreme choice of $|v|=0$ in the co-rotating frame causes
the blue and red regions to extend to the boundary of the dark green
region; all particles outside of the critical ZVC are forbidden to
cross into the area enclosed by the ZVC. Conversely, if the velocities
in the co-rotating frame are large, the red region can vanish and the
blue region can recede far from the binary. This, however, is just a
result of filling the system with initially unbound particles and has
little meaning for an accretion disc. Choosing a purely Keplerian
velocity distribution can also cause issues. The Keplerian velocity
approaches infinity at the origin regardless of primary position. The
result is an artificial depletion of trapped inner-disc particles for
larger mass ratios. Thus, to create a representation that
realistically describes the restricted regions in a quasi-steady-state
CBD, one must choose a velocity distribution near an equilibrium state
(though one may not exist for larger mass ratios) and without
artificial singularities. This leads us to Eq. (\ref{Eq:AddV}).

\begin{figure}
\begin{center}
\includegraphics[scale=0.56]{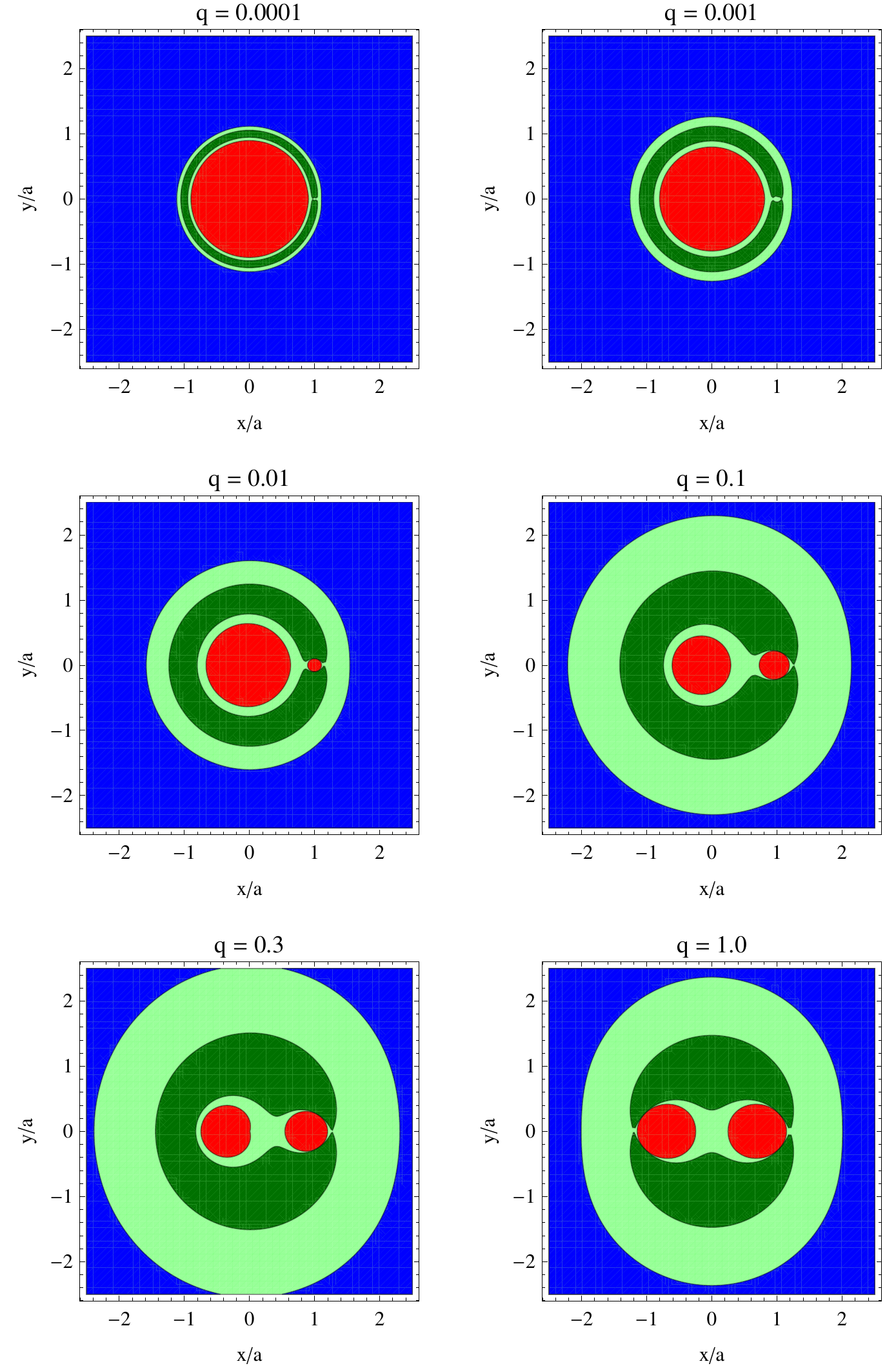}
\end{center}
\caption{The dark green regions are bounded by the zero-velocity curve
  which passes through L2, delineating the smallest restricted regions
  which connect and separating the binary plane into distinct inner
  and outer regions. Particles trapped outside (inside) of the dark
  green region are labeled blue (red). Depending on their velocity
  vectors, light- and dark-green particles are free to move from inner
  to outer regions. }
\label{Fig:CJAddV}
\end{figure}

Because the green regions in Figure \ref{Fig:CJAddV} consists of
particles which are free to cross the binary orbit, we identify these
regions with a putative gap/cavity. We examine this claim in \S
\ref{Integration of the equations of motion} and find that their
locations provide an adequate tracer for the gap/cavity size and shape. 
Furthermore, the locations of the outer gap/cavity edge identified in this manner, and those 
found in Figure \ref{Fig:Int_Orb} below, agree with the 
locations of circumbinary disc truncation computed from the stability 
and intersection of periodic R3Bp orbits \citep{RP:Excretion:1981}. This suggests a correspondence between locations in the outer binary potential where periodic orbits exist, and locations where particles with approximately Keplerian velocity are trapped outside the binary orbital barrier.

We note that the meaning of our Jacobi constant analysis in regions where periodic orbits cannot exist is somewhat ambiguous. In these regions, we must assign velocities to particles which do not correspond to fluid velocities in a stable disc.  \cite{RP:Excretion:1981} have delineated these regions explicitly, and they turn out to coincide with the green, untrapped particles of our analysis. As long as this is true, our analysis is consistent with particle stability considerations.

As the binary mass ratio is increased, Figure
\ref{Fig:CJAddV} shows that the putative gap grows in size and also
morphs in shape from a small horseshoe, or annulus, in the orbit of
the secondary, to a cavernous shape which includes the inner
region. Additionally, as the mass ratio is increased, the fraction of
particles trapped in a disc around the primary decreases while the
number of particles trapped in a disc around the secondary
increases. The following picture appears: for small mass ratios, the
system consists of an inner disc (red) and outer disc (blue) separated
by a low-density annulus. For larger mass ratios, Figure
\ref{Fig:CJAddV} depicts circum-primary and circum-secondary
(mini-)discs (red) surrounded by a CBD (blue). The change in
disc morphology is the most stark between the $q=0.01$ and $q=0.1$
panels of Figure \ref{Fig:CJAddV}.

Within the analysis so far, the transition between an inner and outer
disc, vs. mini-discs+CBD, is defined only by terminology. There is
however an important dynamical transition which occurs within the R3Bp
for mass ratios $q> 0.04$, namely the loss of stable orbits around the
L4/L5 equilibrium points in the binary co-orbital region. In what
follows we provide evidence for a well-defined and physically
meaningful critical mass ratio, related to this stability criterion,
which divides the two regimes described above.

\subsection{Restrictions on Orbits from Equations of Motion}
\label{Integration of the equations of motion}
We now elaborate on the picture painted in Figure \ref{Fig:CJAddV} by
integrating the circular R3Bp equations of motion for a disc of
$256^2$ test particles with initial velocity given by
Eq. (\ref{Eq:AddV}). We paint each of the particles the colour
corresponding to its initial location in Figure
\ref{Fig:CJAddV}. Using an adaptive step, Dormand Prince, 5th order
Runge Kutta method \citep{NumRec:2007}, we evolve the orbit of each
particle for 100 binary orbits, conserving the Jacobi
Constant to fractional order better than $10^{-6}$ (the majority of
orbits conserve $C_J$ to machine precision). Note that, for
large-mass-ratio binaries, a small fraction of the green particle orbits
conserve $C_J$ to only the $10^{-2}$ level. This occurs for green
particles which undergo large accelerations and move to large
distances early in the evolution and does not affect the fate of the
disc.

\begin{figure}
\begin{center}$
\begin{array}{c c }
 \includegraphics[scale=0.38]{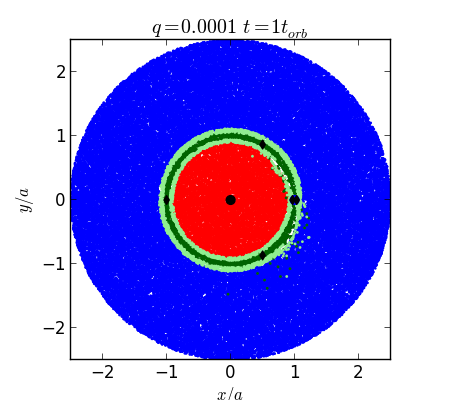} & \hspace{-15 pt}
\includegraphics[scale=0.38]{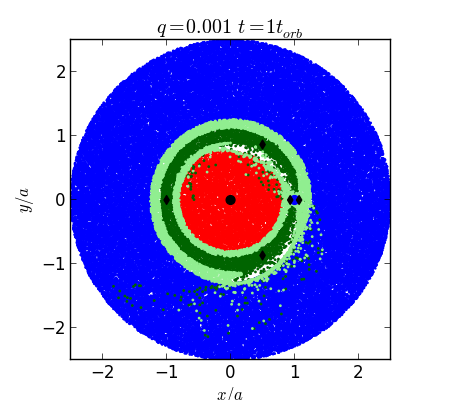} \\
\includegraphics[scale=0.38]{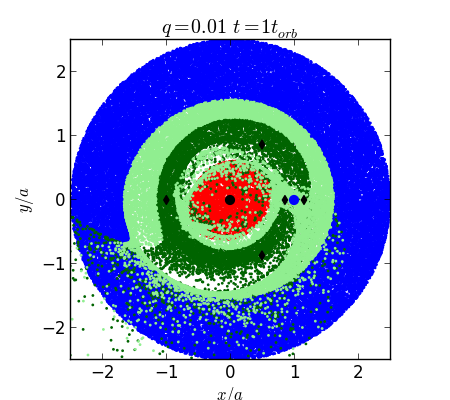} & \hspace{-15 pt}
 \includegraphics[scale=0.38]{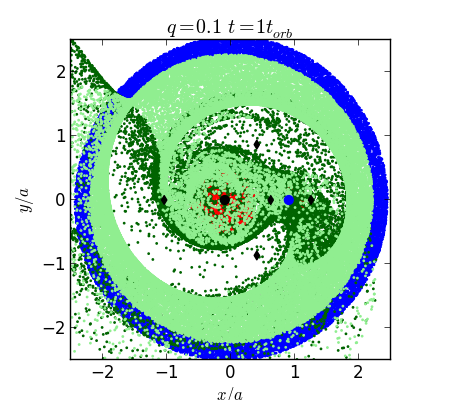} \\
 \includegraphics[scale=0.38]{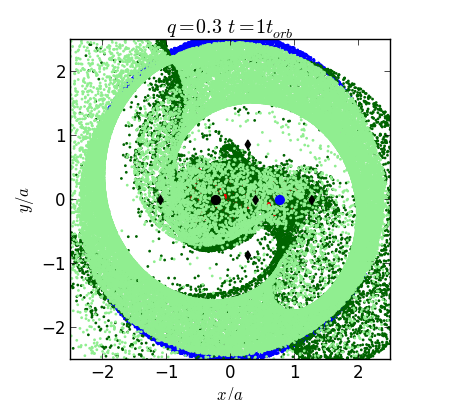} & \hspace{-15 pt}
 \includegraphics[scale=0.38]{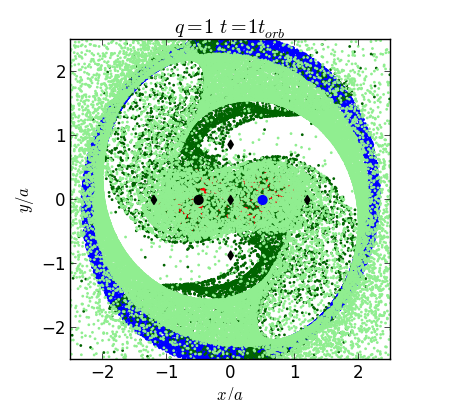}
\end{array}$
\end{center}
\caption{Each panel is the result of evolving an initially spatially random
  distribution of particles, within radius $r<2.5a$, via the R3Bp
  equations, for one binary orbital period. The colouring of particles
  refers to the initial placement of a particle as in Figure
  \ref{Fig:CJAddV}. The black diamonds mark the Lagrange points (see
  Figures \ref{Fig:Roche3D} and \ref{Fig:CJ_Ex}). These snapshots, after only one binary
  orbit, show the formation of streams acting to deplete green particles.}
\label{Fig:Int_Orb_Strms}
\end{figure}

Figure \ref{Fig:Int_Orb_Strms} shows the location of each particle
after only one binary orbit. During the first binary orbit, for mass
ratios $q \gsim 0.04$, the green regions funnel towards the L2 and L3
points into streams reminiscent of those seen in hydrodynamical
simulations and also in the R3Bp study of
\cite{DHM:2013:MNRAS}. Recall that the green particles have Jacobi
Constant corresponding to ZVCs which are not connected, but which
still delineate a restricted region (red-dashed curve in Figure
\ref{Fig:CJ_Ex}). The ZVCs of the green particles allow transfer of the green particles
between inner and outer disc via the lowest barriers in the Roche
potential, the L2 and L3 points respectively (the lowest point in the
potential, L1, allows transfer between primary and secondary - see Figures \ref{Fig:Roche3D} and \ref{Fig:CJ_Ex}). Without
additional forces due to viscosity, pressure, or particle self gravity, the streams do not
persist after a few orbits. If, however, dissipative forces refill the
green region, or particles can interact, the streams continue to form as is seen in
hydrodynamical simulations (see Figure \ref{Fig:Int_Orb_Vsc} of the
Appendix).

For smaller mass ratios $q \lsim 0.04$, the L3 equilibrium point is
much higher than the L2 point and particles stream in horseshoe orbits
past L2 only. The particle density structure, in the $q=0.0001$ and $q=0.001$ cases 
(Figure \ref{Fig:Int_Orb_Strms}), is reminiscent of low-mass-ratio hydrodynamical
simulations where a spiral density wave propagates from the location of the secondary 
(see the top left panel of Figure \ref{Fig:Hydro} below).

Figure \ref{Fig:Int_Orb} shows the final location of each particle,
after 100 binary orbits. We find that the red and blue particles do
indeed remain trapped on their respective sides of the binary
orbit. For $q\leq 0.001$ the green particles move along horseshoe and
tadpole (horse-pole) orbits. As we approach $q=0.01$, we find green
particles librating around the L4 or L5 points. Figure
\ref{Fig:IntOrb_refined} zooms in on mass ratios between $q=0.02$ and
$q=0.08$; for $q\lsim 0.04$, some green particles remain in the
co-orbital region. For mass ratios $q \gsim 0.04$, green particles are
cleared from the binary orbit. This is due to the lack of stable L4/L5
orbits in the R3Bp for $q>0.04$. The widening of the annulus and loss
of particles on horse-pole orbits at $q \simeq 0.04$ marks a
dynamically significant transition caused by a change in orbital
stability. This loss of orbital stability will be especially important
in the hydrodynamical case, where particles can interact.

\begin{figure}
\begin{center}$
\begin{array}{c c }
 \includegraphics[scale=0.38]{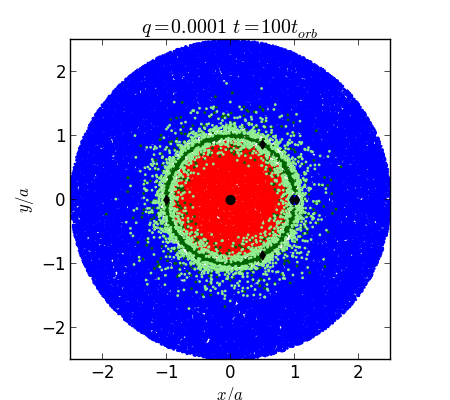} & \hspace{-15 pt}
\includegraphics[scale=0.38]{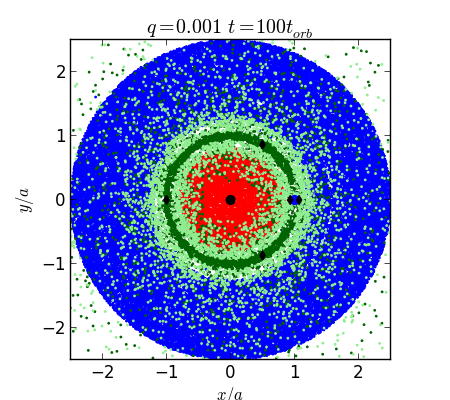} \\
\includegraphics[scale=0.38]{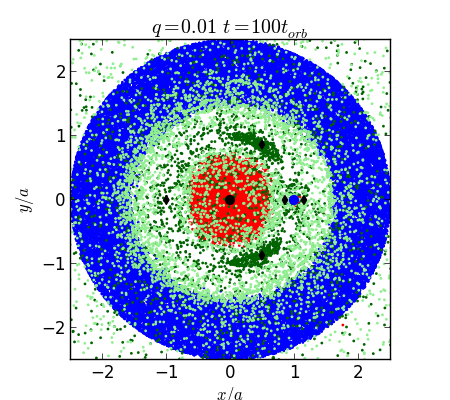} & \hspace{-15 pt}
 \includegraphics[scale=0.38]{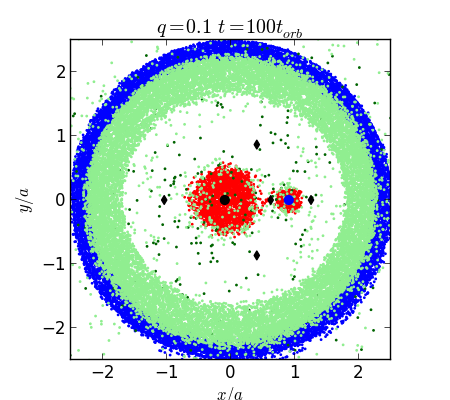} \\
 \includegraphics[scale=0.38]{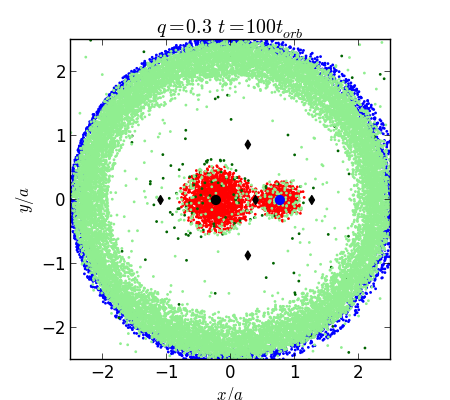} & \hspace{-15 pt}
 \includegraphics[scale=0.38]{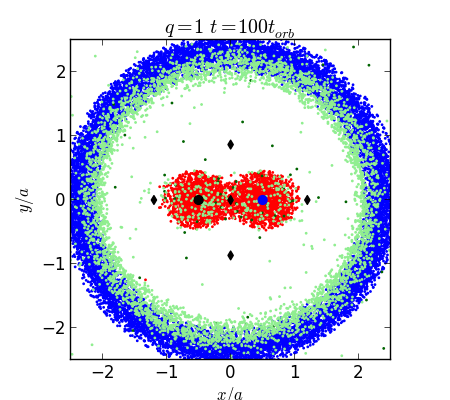} 
\end{array}$
\end{center}
\caption{Same as Figure \ref{Fig:Int_Orb_Strms} except after 100 binary orbital periods.}
\label{Fig:Int_Orb}
\end{figure}

\begin{figure}
\begin{center}$
\begin{array}{c c}
  \includegraphics[scale=0.38]{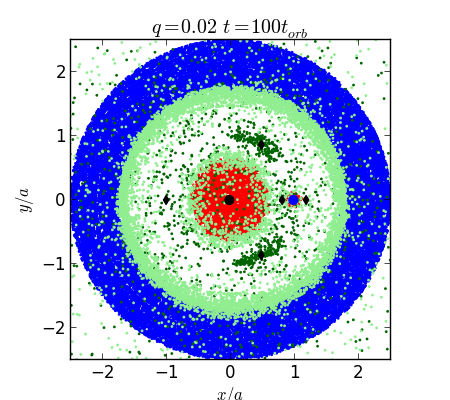} & \hspace{-15 pt}
 \includegraphics[scale=0.38]{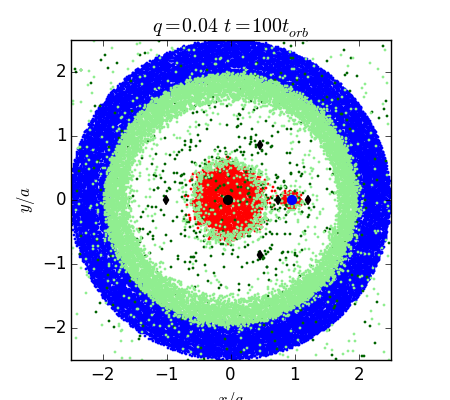} \\
  \includegraphics[scale=0.38]{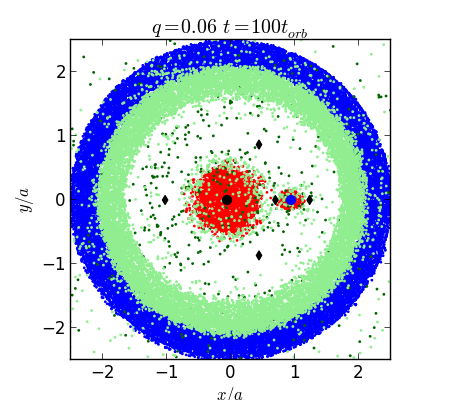} & \hspace{-15 pt}
 \includegraphics[scale=0.38]{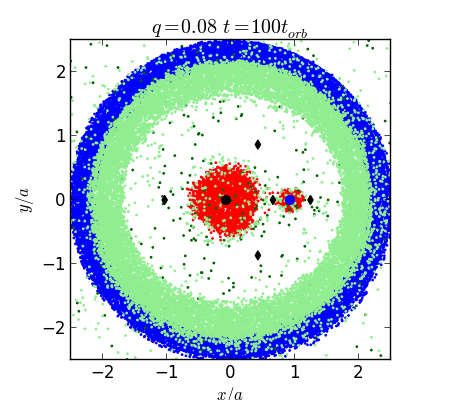} 
\end{array}$
\end{center}
\caption{The same as Figure \ref{Fig:Int_Orb} except zooming in on the
  mass-ratio range $0.02 \leq q \leq 0.08$.}
\label{Fig:IntOrb_refined}
\end{figure}

\subsection{Hydrodynamical Effects}
\label{Viscous and Pressure Effects}
Before numerically solving the viscous hydrodynamical equations, we attempt to estimate the effects of pressure and viscosity by making simple extensions to the standard R3Bp. Pressure and viscosity provide additional forces in the
R3Bp equations of motion which destroy the conservation of $C_J$ and 
modify orbital stability in the standard R3Bp.

\subsubsection{Pressure}
\label{Pressure}
Hydrodynamic pressure can be thought of
as altering the effective potential of the binary at any point along
a particle's trajectory thus altering the ZVCs and hence the
boundaries of the restricted regions in the previous section. 
We estimate the magnitude of pressure below which the R3B analysis may still be relevant by comparing 
the Jacobi constant to the analogue for a pressurized flow, Bernoulli's constant
\begin{equation}
C_B =  2U - v^2 - 2\int{\frac{dP}{\rho}},
\label{Eq:CB}
\end{equation}
where the integral is along the trajectory of a fluid element from a
reference point to the point of evaluation.  Subtracting Eqs.
(\ref{Eq:CJ}) and (\ref{Eq:CB}) we find $2\int{dP/\rho}= C_J - C_B$. 
By setting the difference in integrals of motion equal to the change in 
Jacobi constant across the binary orbit, we estimate the level of disc pressure 
necessary to overflow the previously described restricted regions of the purely gravitational problem \citep[see][who perform a similar calculation]{PR:InnerDsks:1981, RP:Excretion:1981}. For an ideal gas,
\[
\nonumber 
\int{\frac{dP}{\rho}} = \frac{(c^{\rm{ad}}_{s})^2 }{\gamma -1}  = \frac{\gamma (c^{\rm{iso}}_{s})^2 }{\gamma -1} \quad \rm{Adiabatic} \ \rm{Flow} \nonumber
\]
\begin{equation}
\label{dP_iso}
\int{\frac{dP}{\rho}} = (c^{\rm{iso}}_{s})^2 \ln{\frac{\rho}{\rho_0}} \quad \rm{Isothermal} \ \rm{Flow},
\end{equation}
where the isothermal sound speed $c^{\rm{iso}}_{s}$ is related to the
adiabatic sound speed $c^{\rm{ad}}_{s}$ by a factor of the adiabatic
index $\gamma$. In the last line, the ratio $\rho/\rho_0$ comes from
integrating from reference position outside of the binary orbit at
density $\rho_0$, to a point inside the putative gap/cavity at density $\rho$.

The condition for vertical hydrostatic equilibrium in a thin Keplerian
disc is $P/\rho = (GM/r)(H/r)^2$. This allows us to write the sound
speed $\gamma(c^{\rm{iso}}_s)^2 = (c^{\rm{ad}}_s)^2 = \gamma
\Omega^2_K H^2$ in terms of the Keplerian angular frequency of the disc
$\Omega_K$ and disc height $H$. Then the disc orbital Mach number can
be expressed as $\mathcal{M} \equiv v_k/c^{\rm{iso}}_s =
(H/r)^{-1}$, encoding the temperature and pressure forces for a disc in vertical 
hydrostatic equilibrium. From vertical hydrostatic balance, we place a condition on the disc aspect ratio, at the location of the secondary, for which pressure forces can overcome the binary gravitational barrier, 
\[
\nonumber 
\left(\frac{H}{r_s} \right)^{-1}_{\rm{ad}}  \lsim  \sqrt{2 \frac{\gamma}{\gamma-1} \frac{GM_{\bin}}{ \Delta C^{\rm{gap}}_J  a} (1+q) } \nonumber 
\]
\begin{equation}
\left(\frac{H}{r_s} \right)^{-1}_{\rm{iso}}  \lsim  \sqrt{2 \left| \ln{\left( \frac{\rho}{\rho_0}\right)} \right| \frac{GM_{\bin}}{ \Delta C^{\rm{gap}}_J a} (1+q) },
\label{Eq:Mach_Cls}
\end{equation}
where $\Delta C^{\rm{gap}}$ is the variation of $C_J$ across the
dark-green restricted regions of Figure
\ref{Fig:CJAddV}. Operationally, we choose $\Delta C^{\rm{gap}}$ to be
the difference in $C_J$ at L2 and L4 (or L5), as this is the largest
$\Delta C$ spanning the dark-green restricted regions.

We emphasize that Eqs. (\ref{Eq:Mach_Cls}) are not gap/cavity closing conditions; 
they are necessary conditions for the pressure to overcome the gravitational barrier of the binary. These conditions do not take into account the direction of pressure forces, and they do not consider hydrodynamical shocks, which have been shown, in the small mass ratio regime ($q\leq 10^{-3}$), to be responsible for gap opening in competition with viscous forces \citep{DongRafI:2011, DongRafII:2011, DuffellMac:2013:smallqGapOpen, FungGaps:2014, DuffellGapAnyl:2015}. 
True gap/cavity closing conditions must incorporate a full hydrodynamical treatment, whereas the conditions (\ref{Eq:Mach_Cls}) provide a necessary condition for pressure to dominate the flow dynamics.

\begin{figure}
\begin{center}
\includegraphics[scale=0.25]{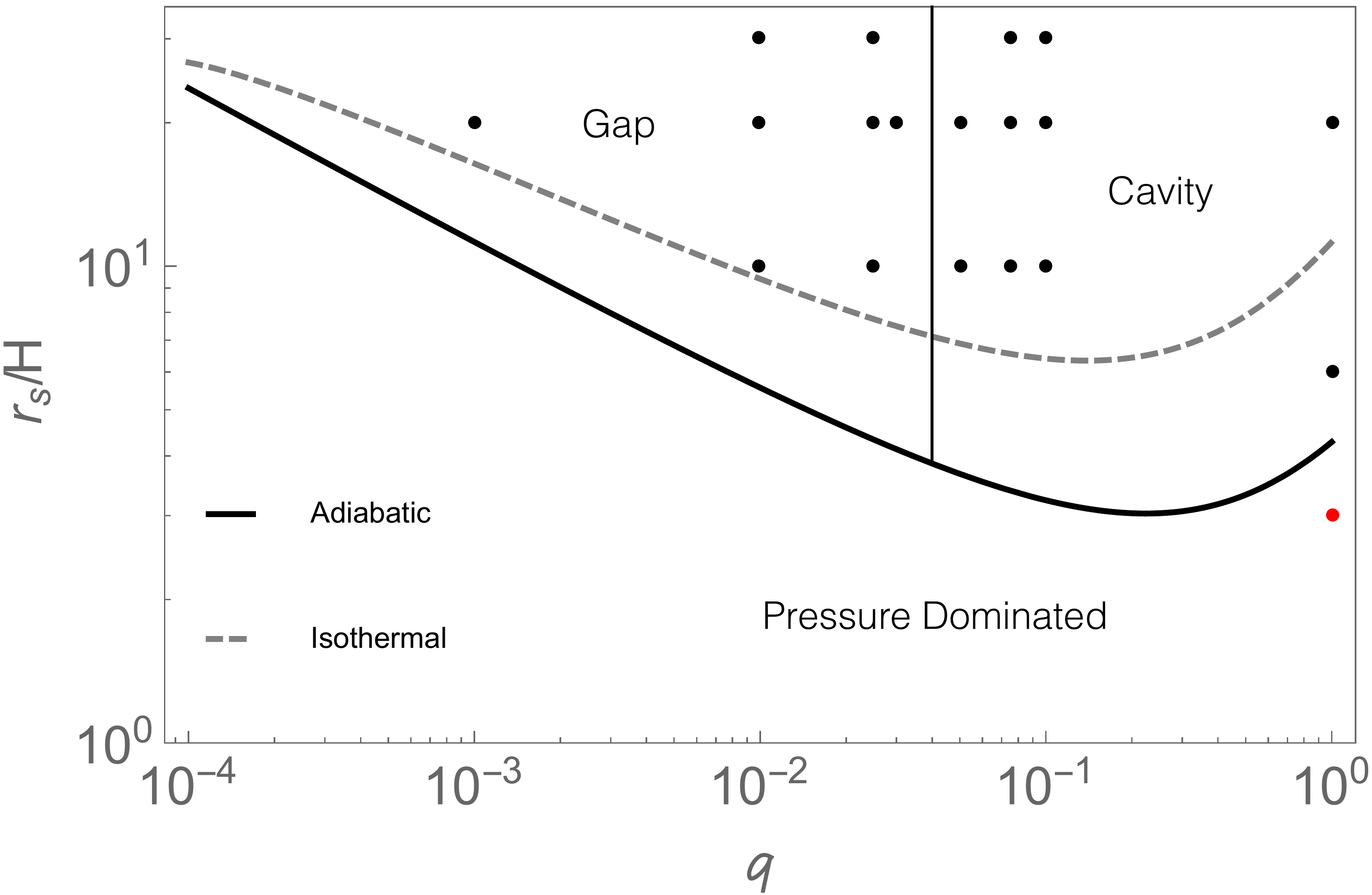} \vspace{-10 pt}
\end{center}
\caption{Delineation of different phases in a non-viscous circumbinary disc. The y-axis 
  records the inverse disc aspect ratio, equivalent to the orbital Mach number; a 
  smaller value signifies larger pressure forces; large pressure forces preclude 
  reasoning based on a purely gravitational analysis. Points
  represent the parameters of hydrodynamical simulations run in this study 
  (\S \ref{Hydrodynamical Simulations}). Red denotes a simulation with 
  a filled gap/cavity.}
\label{Fig:GapOpen}
\end{figure}

\begin{figure}
\begin{center}$
\begin{array}{cc}
\hspace{-16 pt}
\includegraphics[scale=0.24]{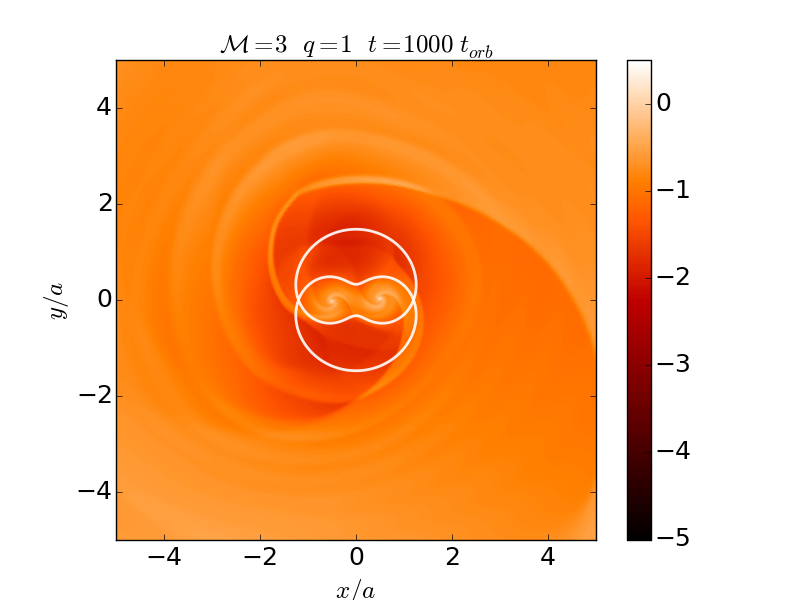} &
\hspace{-20 pt}
\includegraphics[scale=0.24]{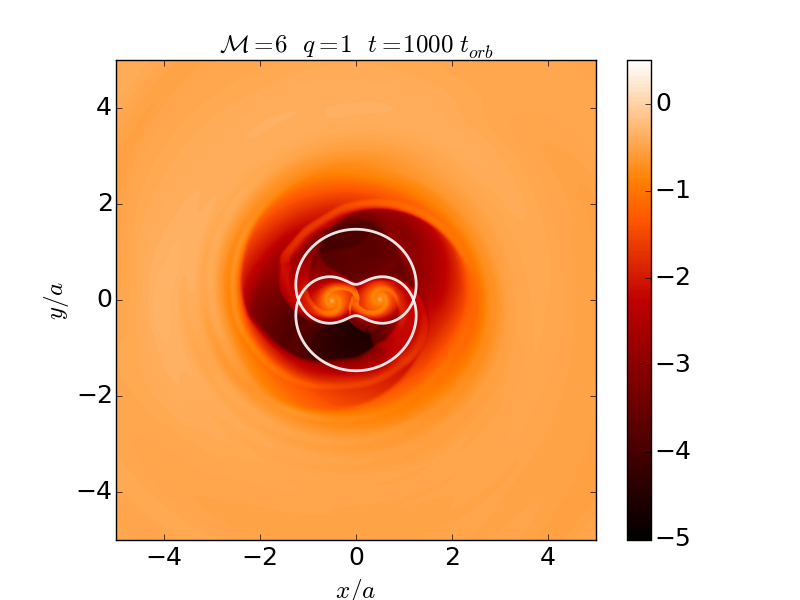} \\
\hspace{-16 pt}
\includegraphics[scale=0.24]{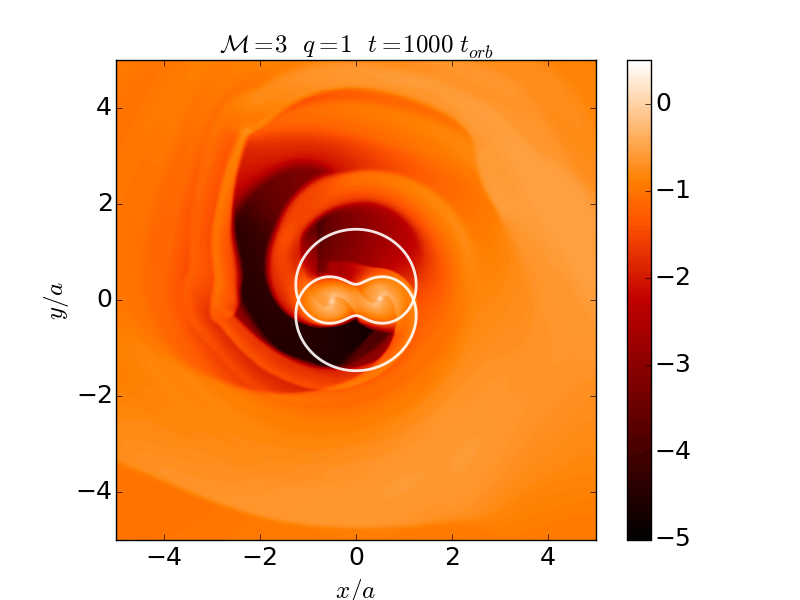} &
\hspace{-20 pt}
\includegraphics[scale=0.24]{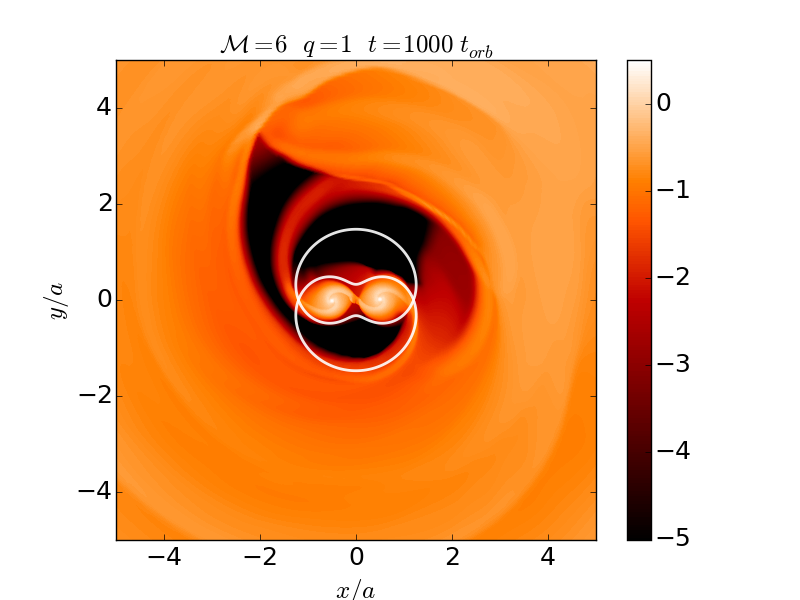}  
\vspace{-10 pt}
\end{array}$
\end{center}
\caption{Snapshots of the surface density distribution (shown in units
  of the unperturbed value, with a logarithmic colour scheme) for an
  equal mass binary with disc aspect ratios ($r_s/H \equiv \mathcal{M}$) surrounding 
  the pressure dominated condition Eq. (\ref{Eq:Mach_Cls}). 
  Here we set the viscosity to be very
  small (the coefficient of kinematic viscosity is $\nu =
  10^{-6}a^2_0\Omega_{\bin}$, where $\Omega_{\bin}$ is the binary
  angular frequency) in order to examine the analytic R3Bp
  prediction (\ref{Eq:Mach_Cls}). The top row is for an isothermal equation of state
  $P=(c^{\mbox{iso}}_s)^2 \Sigma$ and the bottom row is for an
  adiabatic equation of state $P=(c^{\mbox{ad}}_s)^2 \Sigma^{5/3}$.}
\label{Fig:Gap_Open_q1Sims}
\end{figure}

In Figure \ref{Fig:GapOpen} we plot critical disc aspect ratios
(\ref{Eq:Mach_Cls}) as a function of binary mass ratio. Over-plotted dots mark
the positions of hydrodynamical simulations run in this study (\S \ref{Hydrodynamical Simulations}). For the density ratio
$\rho/\rho_0$ across a gap/cavity, we use the empirical relation from
\cite{DuffellMac:2013:smallqGapOpen}. All of the primary simulations run in 
this study have disc aspect ratios below the critical limit. 
To probe this limit we run high aspect ratio, $q=1$ hydrodynamical simulations for the
isothermal and adiabatic cases (see \S \ref{Hydrodynamical Simulations}). 
Figure \ref{Fig:Gap_Open_q1Sims} shows that the results of
these high-aspect-ratio (equivalently, high pressure, thick disc, high temperature,
or low-orbital-Mach number) simulations\footnote{In these hot discs, we often see a one-armed spiral structure similar to that reported by \cite{ShiKrolik:2015}, also for hotter discs.} are in agreement with our
Eqs. (\ref{Eq:Mach_Cls}). The isothermal case shows a stark difference between an
overflowed cavity at $H/r_s=1/3$ and a cleared cavity at $H/r_s=1/6$ while the 
adiabatic case is less extreme, with the $\Mach=3$ adiabatic case exhibiting a marginal central cavity.

\cite{DuffellMac:2013:smallqGapOpen} have shown that in the small mass ratio, thin disc case, the clearing of a gap always occurs for inviscid discs. This is because, in the absence of viscous forces, density waves generated by the binary will shock and deposit angular momentum into the disc clearing a gap without competition. It is not clear wether this is always true for high mass ratio, highly-pressurized discs; the overflowed CBDs in Figure \ref{Fig:Gap_Open_q1Sims} may eventually clear a cavity due to shocks, overcoming pressure on a longer timescale than considered here. 

In summary, a necessary condition for pressure forces to overflow 
the binary cavity is predicted by Eq. (\ref{Eq:Mach_Cls}). Equal-mass binary 
simulations show this condition to be sufficient in the large 
mass ratio regime, at least initially. Hence, having a high pressure in 
the disc does not impede the cavity formation, or whether it is lopsided,
unless the pressure becomes so large that the disc is no longer thin. 
In this case, 3D effects can also become important, invalidating the 2D analysis above.

\subsubsection{Viscosity}
\label{viscosity}
In order to more closely compare to viscous hydrodynamical
simulations, and with the goal of linking orbital stability to the CBD
phase transition, we follow \cite{Murray:1994} and \cite{MD:SSD} to
add an external viscous force $\mathbf{F}(x,y,\dot{x},\dot{y})$ to the
R3Bp equations of motion. With these viscous R3Bp equations, we conduct
a linear-stability analysis and integrate the viscous R3Bp equations
of motion for a disc of test particles. We refer the interested reader
to the Appendix for details; here we only state a brief summary of the
three main conclusions.
\begin{enumerate}
\item Upon integrating the viscous R3Bp equations for an initial disc
  of test particles, we find that viscosity indeed acts to overflow
  gaps in the R3Bp and continually generates streams which penetrate
  the binary orbit (compare Figure \ref{Fig:Int_Orb_Vsc} to Figure
  \ref{Fig:Int_Orb}). The time rate of change of the Jacobi constant
  due to viscosity is given for a simple Keplerian velocity
  prescription in the Appendix.
\item The addition of viscosity causes orbits around L4 and L5 to
  become formally unstable for all binary mass ratios (Figure
  \ref{Fig:Lam}). However, the instability timescale for orbits below
  $q=0.04$ is of order a viscous time, while for $q>0.04$ it drops to
  of order the binary orbital time. Hence, viscosity does not greatly
  change the mass ratio at which orbits around L4 and L5 become
  effectively unstable. If the phase transition at $q=0.04$ is linked
  to the instability of orbits in the co-orbital region of the binary,
  then the level of viscosity should not greatly affect the mass ratio
  of the $q=0.04$ phase transition - for thin discs for which
  Eqs. (\ref{Eq:Mach_Cls}) hold. We verify this with a suite of
  hydrodynamical simulations in the next section.
\item Viscosity induces a difference in orbital instability timescales
  for particle orbits around the L4 and L5 points (Figure
  \ref{Fig:Lam_L4L5}). This difference is proportional to the magnitude 
  of viscosity and it is small for $q<0.04$, becoming larger for $q>0.04$. 
  While this asymmetry between L4 and L5 may aid in seeding the 
  instability to an asymmetric cavity, it cannot be the only mechanism 
  which causes the cavity to be lopsided.  For example, symmetry 
  between L4 and L5 must be restored at $q=1$, and a prominent asymmetry 
  still appears in the disc morphology in this case.  Future work will 
  explore in more detail the relationship between the different orbital 
  stability timescales and the cavity morphology.
\end{enumerate}

\section{Hydrodynamical simulations}
\label{Hydrodynamical Simulations}
\subsection{Fiducial Simulations}
\label{Fiducial Simulations}
In this section we show that the intuition gained from the R3Bp
carries over to the hydrodynamical regime by running viscous 2D
hydrodynamical simulations of the binary-disc system.  We utilize the
moving mesh code DISCO \citep{Duffell:2011:TESS} to simulate a binary
embedded in a locally isothermal, initially uniform surface density
($\Sigma = \rm{cst.}$) disc. To enforce a locally isothermal equation
of state, the vertically integrated pressure is set as
$P=(v_{\rm{eff}} / \Mach)^2 \Sigma$, where
\begin{equation}
v_{\rm{eff}} =  \sqrt{\frac{GM_s}{r_s} + \frac{GM_p}{r_p}}
\end{equation}
is the initial azimuthal velocity in the disc for an initially,
spatially constant Mach number $\Mach$, and $\Sigma$ is the disc
surface density. The initial radial velocity is given by viscous
diffusion
\begin{equation}
v_r = - \frac{3}{\Sigma r^{1/2}} \partial_{r}\left[ \nu \Sigma  r^{1/2}\right],
\end{equation}
where we choose a coefficient of kinematic viscosity which is constant
in space and time $\nu = \alpha a^2_0 \Omega_{0} /\Mach^2$, for
fiducial values $\alpha = 0.01$ and $\Mach = 20$. These choices assure
gap/cavity formation for $q\gsim10^{-4}$. Here $a_0$ and $\Omega_0$
represent the fixed separation and angular frequency of the
binary. The simulation domain extends from the origin to $r_{\rm{max}} = 12.0a$
employing a log grid ($r_{\rm{max}}$ can range from $8.0a$ to $100.0a$
for non-fiducial disc parameters as discussed in \S
\ref{ParamStudy}). The radial resolution is $\Delta r \sim 0.02a$
inside the binary orbit, similar to that of \cite{Farris:2014}, except
we do not add an additional high-resolution region around each BH. We
choose an outflow outer boundary condition. We do not apply an inner
boundary condition and instead allow gas to flow through six cells at
$r<0.05a$. Around each binary component we employ a density sink of
size $r_{\rm{acc}}$ which removes gas at the rate $3/2 \Sigma r^{-2}_n
\nu$ from each cell within the sink radius, where $r_n$ is the
distance from the $n^{\rm{th}}$ binary component. The total accretion
rate onto each binary component is found by integrating over each cell
located inside of the sink radius. Unless otherwise specified, we set
$r_{\rm{acc}} = \rm{min}\left\{ R_{\rm{Hill}}, 0.5a \right\}$ for the
secondary and $r_{\rm{acc}}= 0.5a$ for the primary, where
$R_{\rm{Hill}}$ is the Hill radius of the secondary.

\begin{figure*}
\begin{center}$
\begin{array}{c c c}
\includegraphics[scale=0.31]{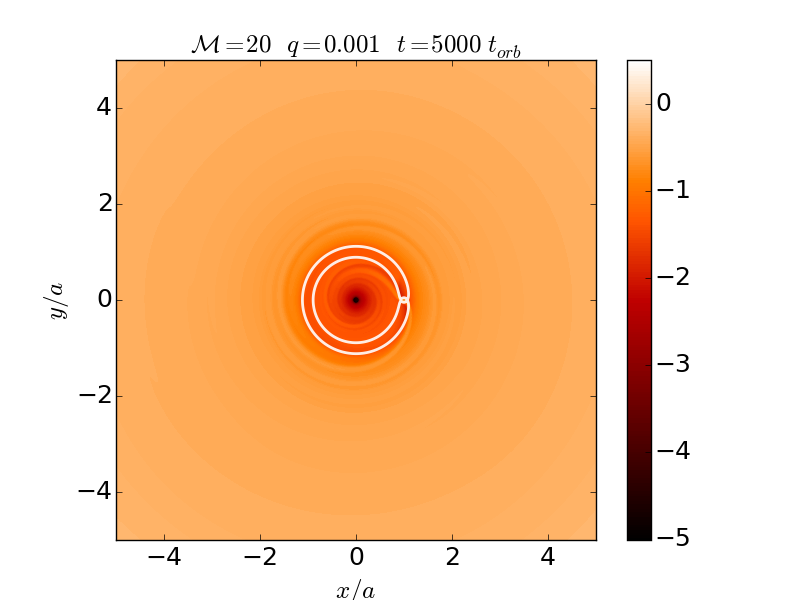}& \hspace{-20 pt}
 \includegraphics[scale=0.31]{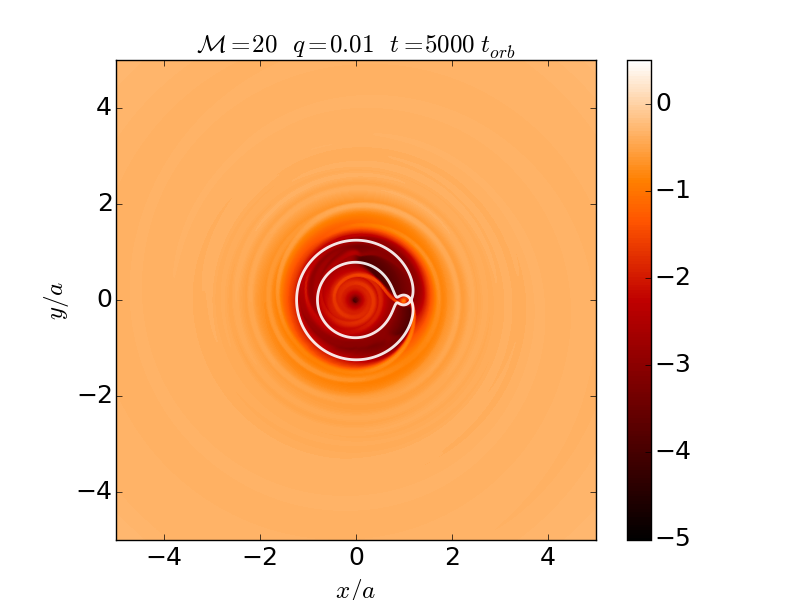} &  \hspace{-20 pt}
 \includegraphics[scale=0.31]{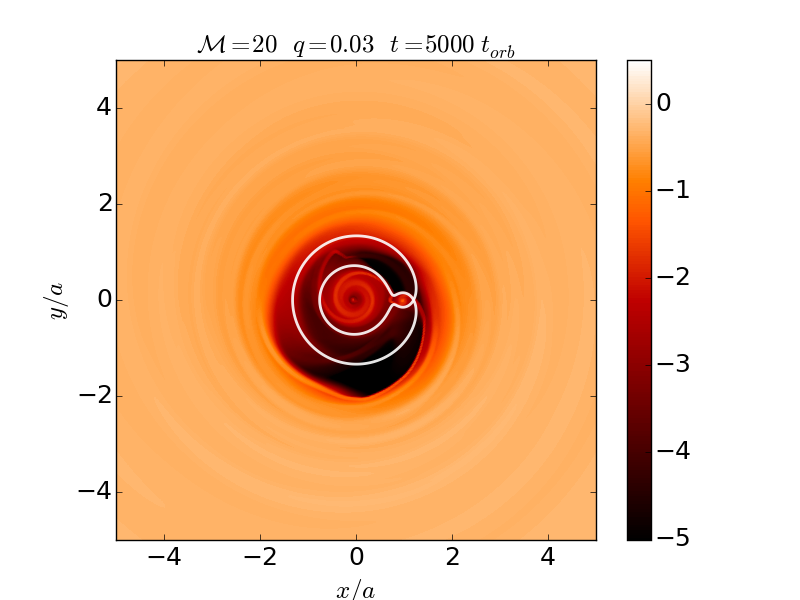} \\
 \includegraphics[scale=0.31]{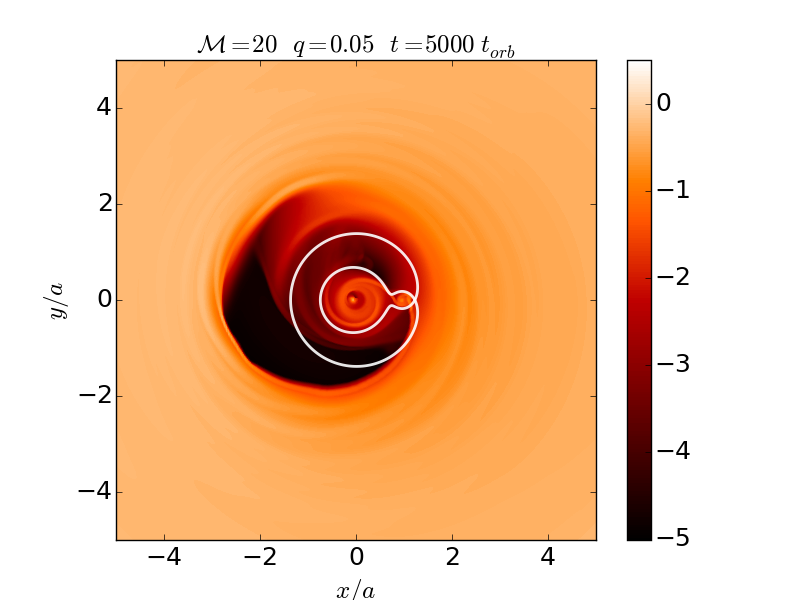} &  \hspace{-20 pt}
\includegraphics[scale=0.31]{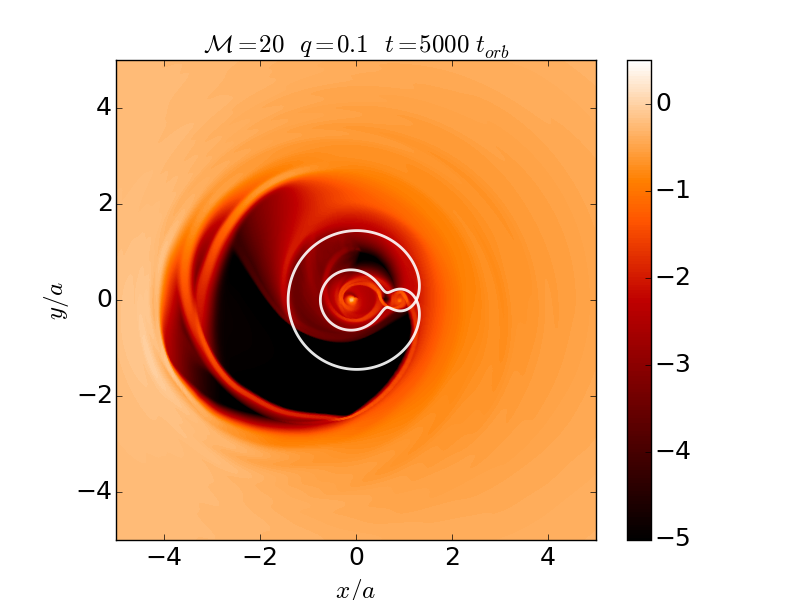} &   \hspace{-20 pt}
\includegraphics[scale=0.31]{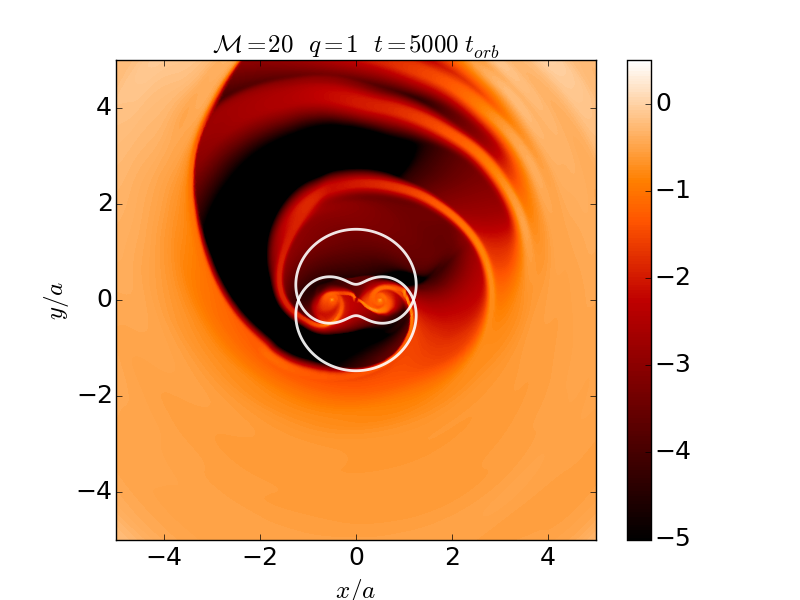}
\end{array}$
\end{center}
\caption{Snapshots of the surface density distribution (shown in units
  of the unperturbed value, with a logarithmic colour scheme) from
  hydrodynamical simulations for a disc with orbital Mach number
  $\Mach \equiv r/H \equiv v_{\rm{eff}}/c_s= 20$, and constant
  coefficient of kinematic viscosity $\nu = 0.01 a^2_0 \Omega_{\bin}
  /\Mach^2$. The binary mass ratio increases from left to right, top
  to bottom, as labeled. For small mass ratios, the system is in nearly
  steady-state and an annular gap is cleared in the orbit of the
  secondary black hole. For $q\gsim 0.03$, the gap morphs into an even
  lower density time-dependent, precessing central cavity. The
  critical zero-velocity curve, which passes through L2, is over-drawn
  in white. The relatively shallow annular gap in the $q=0.001$ case is difficult to see on this scale 
  because the accretion prescription and inner boundary cause the inner disc to 
  drain onto the primary.}
\label{Fig:Hydro}
\end{figure*}

We evolve the viscous hydrodynamical equations for at least one
viscous time at the location of the binary. We plot the resulting 2D surface-density distribution for different mass ratios in
Figure \ref{Fig:Hydro}. We make three observations on the CBD-density structure:
\begin{itemize}
\item \textbf{An annular gap morphs into a central cavity:} Comparing the
  $q=0.03$ and the $q=0.05$ panels in Figure \ref{Fig:Hydro}, we see
  that there is a transition occurring by $q=0.03$ from an annular gap
  to a lopsided cavity (consistent with \cite{Farris:2014} which used
  a hotter and more viscous disc) as well as a constant $\alpha$ (as
  opposed to constant $\nu$) viscosity prescription. We explore the
  dependence of the critical mass ratio on disc Mach number and
  viscosity in \S \ref{ParamStudy} below.
\item \textbf{The surface density decreases in the co-orbital region:} The
  surface density in the annular gap decreases with increasing mass
  ratio until the gap becomes nearly devoid of gas above the transition
  to a lopsided cavity. While there are multiple factors that determine 
  gap depth \citep[and width;][]{FungGaps:2014, Kanagawa:2015, DuffellGapAnyl:2015}, 
  we note that the loss of stable orbits librating around L4/L5 and the disappearance of gas near the orbit of 
  the secondary occur near the same mass ratio, suggesting that the 
  existence of stable L4/L5 orbits may be necessary for long-lived gas in the gap. 
  This behaviour helps to define the phase transition occurring at the same mass ratio.
\item \textbf{A lopsided, precessing cavity appears:} The
  hydrodynamical study introduces a phenomenon accompanying the
  annular-gap to central-cavity transition which is not captured by
  the R3Bp alone, a lopsided, precessing cavity. This is illustrated
  by the surface density contours in the bottom panels of Figure
  \ref{Fig:Hydro}. A lopsided cavity has been observed in previous
  studies which have used different numerical codes and physical
  assumptions \citep{MacFadyen:2008, ShiKrolik:2012:ApJ, Noble+2012, 
  DHM:2013:MNRAS, Farris:2014,Farris:2015:Cool,Farris:2015:GW, 
  Dunhill:2015, ShiKrolik:2015}. A complete description of the growth 
  of lopsided cavities, especially in the symmetric equal-mass binary case, 
  does not yet exist. However, the growth has been attributed to an initially small asymmetry in stream
  generation which causes a feedback loop between stream strength and
  cavity-edge location \citep{ShiKrolik:2012:ApJ, DHM:2013:MNRAS}.  We
  observe that lopsided cavities are generated when stable orbits do
  not exist in the co-orbital region. Hence, the generation of a
  lopsided cavity may be intimately tied to orbital dynamics in the
  co-orbital region. When there are no stable co-orbital orbits, fluid
  passing into the co-orbital regime is flung into the outer disc
  rather than librating on stable orbits.
\end{itemize}

\begin{figure*}
\begin{center}$
\begin{array}{cccc}  
 \includegraphics[scale=0.3]{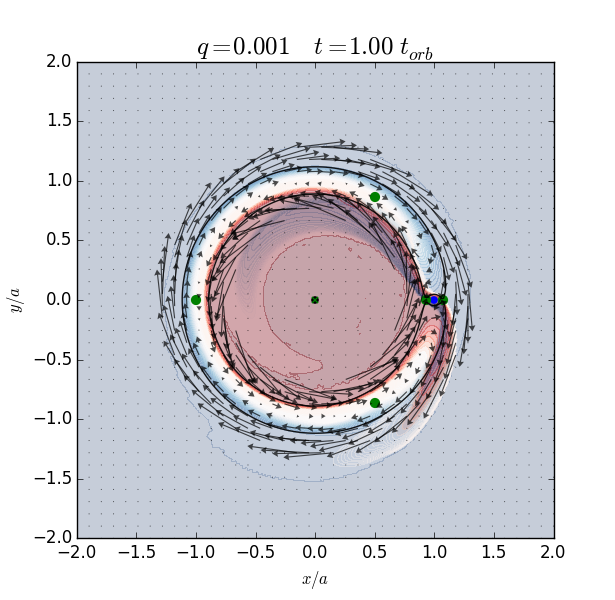} &  \hspace{-20 pt}
 \includegraphics[scale=0.3]{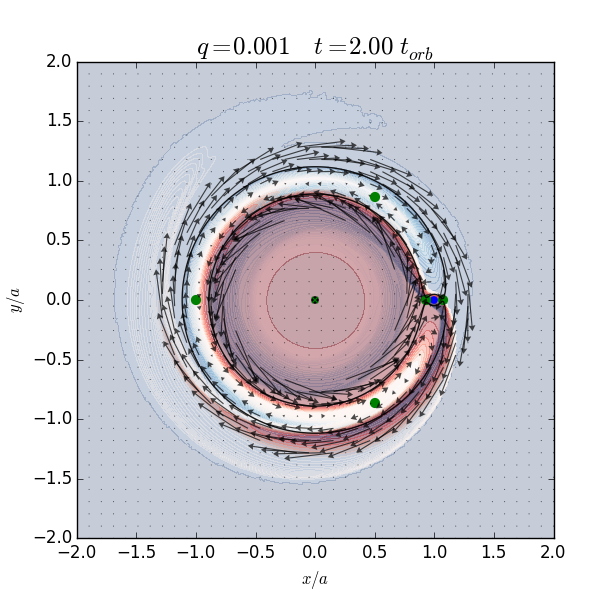} &   \hspace{-20 pt}
 \includegraphics[scale=0.3]{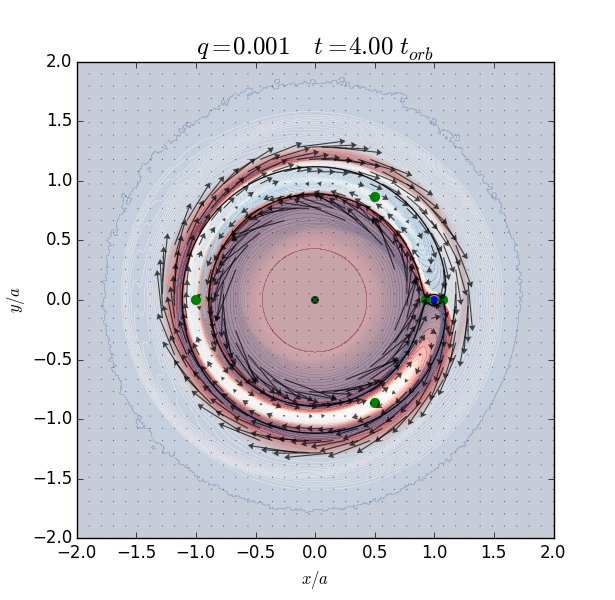} &   \hspace{-20 pt}
 \includegraphics[scale=0.3]{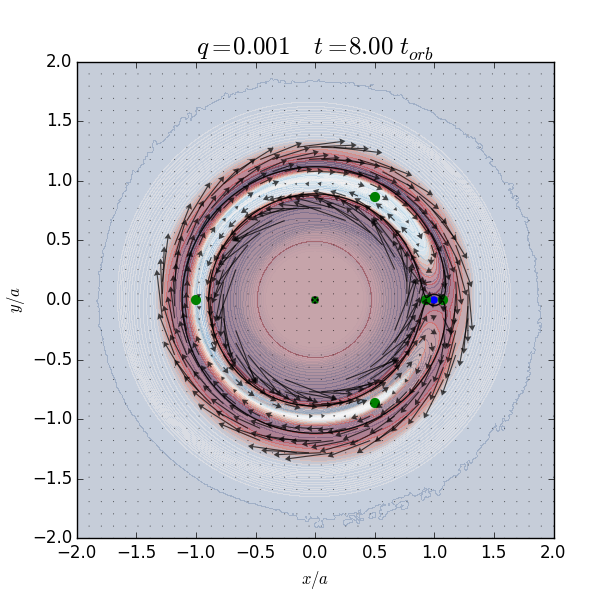}\\  
 \includegraphics[scale=0.3]{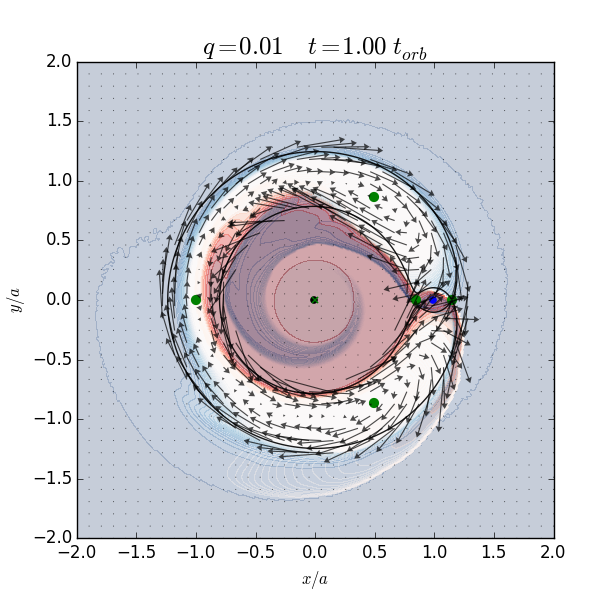} &  \hspace{-20 pt}
 \includegraphics[scale=0.3]{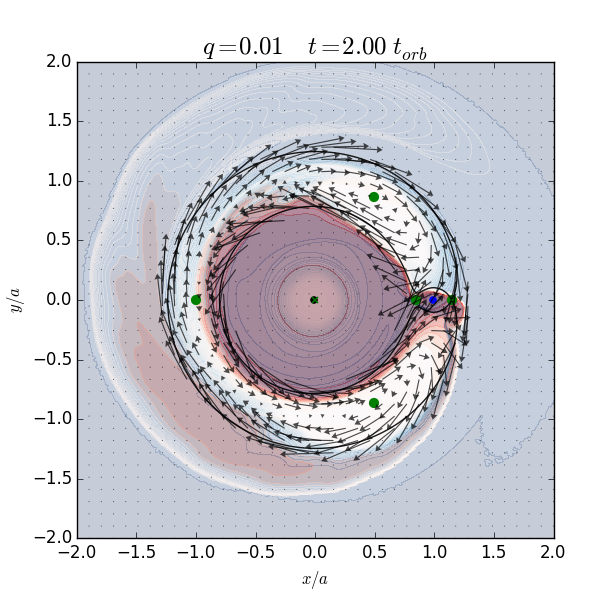} &  \hspace{-20 pt}
 \includegraphics[scale=0.3]{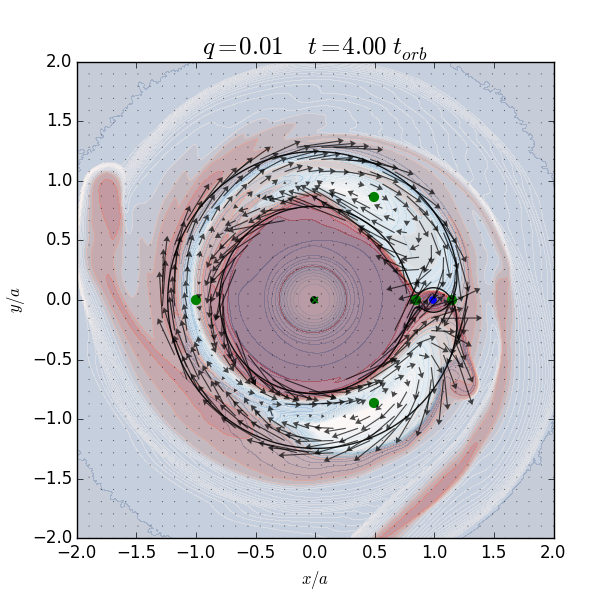} &  \hspace{-20 pt}
 \includegraphics[scale=0.3]{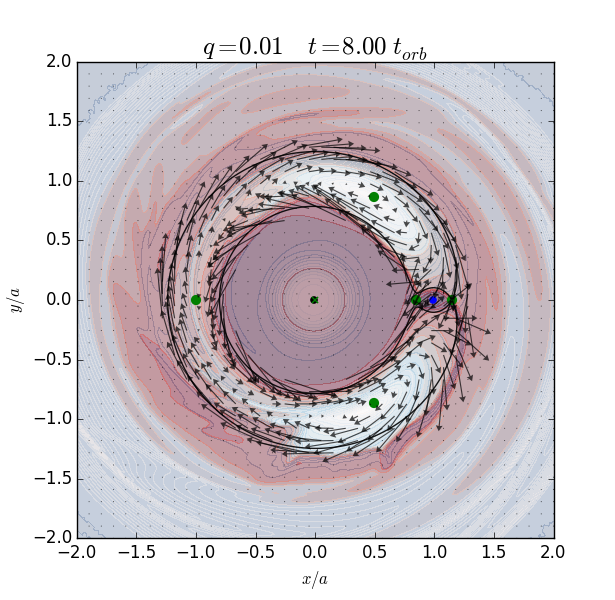} \\   
 \includegraphics[scale=0.3]{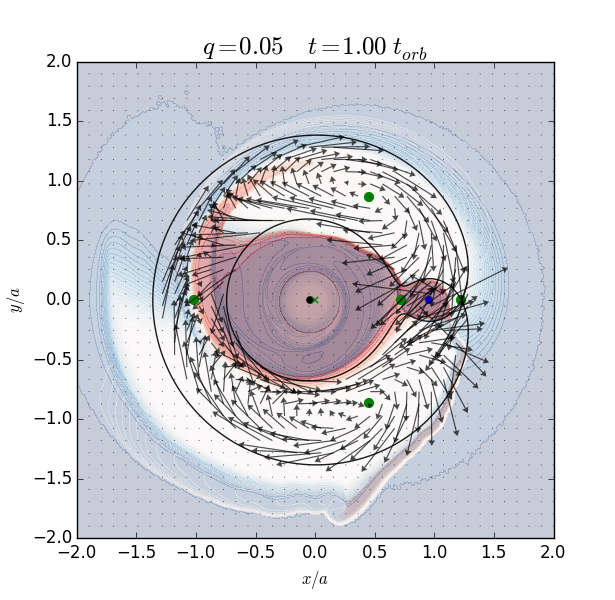} &  \hspace{-20 pt}
 \includegraphics[scale=0.3]{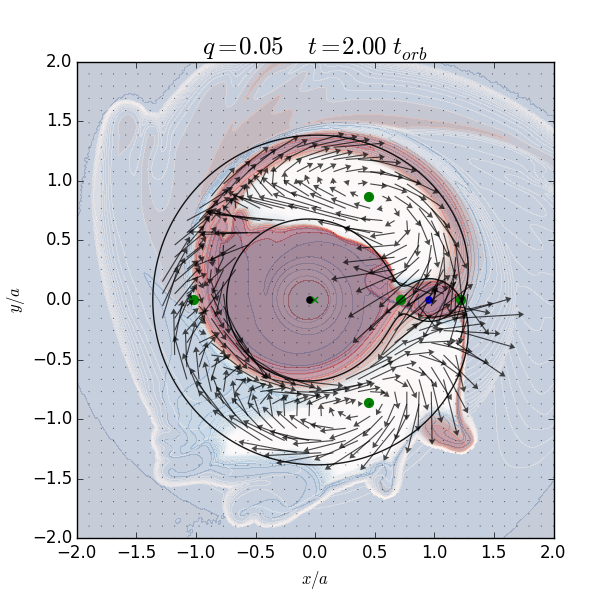} &  \hspace{-20 pt}
 \includegraphics[scale=0.3]{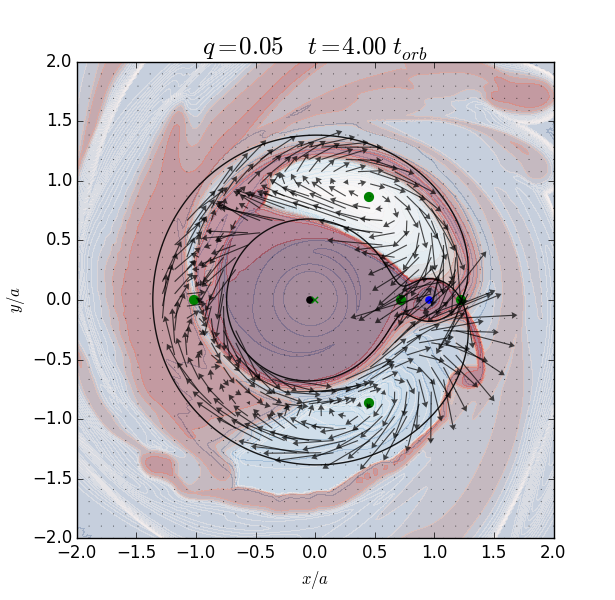} &  \hspace{-20 pt}
 \includegraphics[scale=0.3]{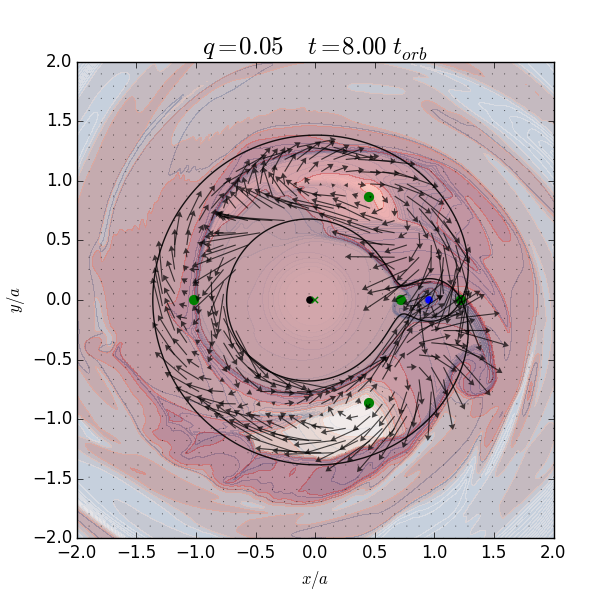}
\end{array}$
\end{center}
\caption{Evolution of two passive scalars for $q=0.001$, $q=0.01$, and
  $q=0.05$ (rows) at different times (columns) during gap opening as
  well as velocity vectors showing fluid motion in the co-rotating
  frame. The red scalar starts inside of the critical ZVC (overlaid,
  black curve) and the blue scalar starts outside. The green dots
  denote the Lagrange points of the binary potential (See Figures
  \ref{Fig:Roche3D} and \ref{Fig:CJ_Ex}). Published with this article
  are {\em movies} of the above three simulations with the red and blue
  passive scalars plotted on different panels and with the same 
  initialization as Figure \ref{Fig:CJAddV} (these movies and the 
  corresponding movies of surface density evolution can also be 
  found at http://user.astro.columbia.edu/$\sim$dorazio/moviespage).}
\label{Fig:PS}
\end{figure*}

To further examine the connection between orbital stability and the
transition to a time-dependent, lopsided cavity, we track the flow of
gas in the co-orbital region. To do this we evolve separate
conservation equations for two passive scalars. A passive scalar is a
scalar quantity which obeys a conservation equation given an initial
concentration and the fluid velocities from the hydrodynamic
problem. We start one scalar outside of the critical ZVC and the other inside. The
light green regions of Figures \ref{Fig:CJAddV} and \ref{Fig:Int_Orb}
show that, even in the non-hydrodynamic case, this setup should result
in the passive scalars moving across the orbit of the binary, which is
what we wish to track. Figure \ref{Fig:PS} plots the evolution of the
passive scalars as well as fluid-velocity vectors for three different
mass ratios. We colour the passive scalar which is initially inside
(outside) of the critical ZVC red (blue).

The first row in Figure \ref{Fig:PS} tracks the fluid motion for a binary 
mass ratio well below the stability/phase transition, $q=0.001$. It is
clear that gas inside the orbit moves outwards across the position of
the secondary at L2 and travels along the horseshoe orbits delineated
by the critical ZVC. Gas initially outside of the orbit similarly
moves across L2 and enters onto horseshoe orbits which eventually
deposit the gas onto the inner disc.

The second row in Figure \ref{Fig:PS} tracks the fluid motion for a
binary with $q=0.01$, which is still below the
R3Bp linear-stability mass ratio. From the velocity vectors it is
clear that the mean motion of the fluid is along the horseshoe and
librating L4/L5 orbits. In the $q=0.01$ case, the velocity of the
fluid is larger in the presence of larger binary forces. The result is
a greater deviation of the gas from the critical ZVC curve, some of
which begins to peel off into the outer disc. As the last panel for
the $q=0.01$ case hints, this behaviour is not sustained once the gap
is cleared and a steady state ensues.

The third row in Figure \ref{Fig:PS} tracks the fluid motion for
$q=0.05$, which is above the mass ratio for which stable L4/L5 orbits
exist. We now see that the red gas immediately flows out along L3 and
L2. The gas leaving L3 connects back to the orbital flow around L4
while the gas leaving L2 is flung into the disc.  A striking feature
is the large velocity vectors pointing outward from between L2 and L5
into the disc rather than pointing back along the horseshoe
orbit. Because of the vigorous ejection of fluid from the L5 point, an
asymmetry builds between the L4 and L5 regions. 

In addition to the ejection of co-orbital particles for $q\gsim 0.04$,
we observe the beginnings of a second stream connecting to the primary
through the L3 equilibrium point, the lowest point in the
binary potential after L1 and L2. The bottom panels of Figure
\ref{Fig:Hydro} show us that, for near equal mass binaries, two accretion streams feed the binary. 
Previous work \citep{ShiKrolik:2012:ApJ, DHM:2013:MNRAS} has argued 
that streams crashing into the surrounding CBD causes lopsided cavity growth. Specifically, \citep{DHM:2013:MNRAS} conducted an experiment where a disc around an equal mass binary is simulated with one binary component placed artificially at the origin of coordinates. In this experiment, only one stream is generated and lopsided growth is inhibited. Hence, though it may not be necessary for lopsided growth, the generation of a second, strong stream for $q \gsim 0.04$ seems to facilitate such growth.

Because the new stream through L3 feeds the primary,
the $q=0.04$ transition also signifies an increase in the accretion
rate of the primary relative to the secondary for $q\gsim0.04$.

Finally, we note that, in order to show the dynamics of gap clearing, 
the flows depicted in Figure \ref{Fig:PS} are chosen at early times in the disc 
evolution. While initial conditions are chosen so that initial transients 
are minimal, such transients may be present, and one should take 
this example only as suggestive of the properties of the flow at later times.

\subsection{Hydrodynamic Parameter Study}
\label{ParamStudy}
Although we identified a clear transition in the disc behaviour near
$q=0.04$, it is natural to ask whether this critical value is universal,
or if it depends on disc parameters. To determine this, we repeat our earlier
hydrodynamical simulations but for two new values of the coefficient
of kinematic viscosity $\nu_0 = 10^{-3} $, $10^{-4}$, where $\nu_0
\equiv \nu a^{-2}_0 \Omega^{-1}_{\bin}$, and two new values of the
Mach number $\Mach = 10$, $30$, corresponding to a factor of 9
variation in disc pressure and temperature. For each of the four new parings of Mach
number and viscosity coefficient, we run four simulations at the
binary mass ratios $q=0.01, 0.025, 0.075, 0.1$.

Note that we increase the outer boundary of the simulation domain to
$r_{\rm max}=16a$ for the $\nu_0 = 10^{-4}$, $\Mach=10$ simulation,
because in these simulations, spiral density waves have longer
wavelengths and are less damped by viscosity, allowing them to reach
the outer boundary of the simulation, with the results therefore
potentially depending on the outer boundary condition. For all other
simulations we choose $r_{\rm max}=8a$. Because we are using a log
grid, this corresponds to only a minimal change in resolution. We have
run higher resolution simulations for the $\nu_0 = 10^{-4}$,
$\Mach=10$ and $\nu_0 = 10^{-3}$, $\Mach=30$ cases with outer boundary
at $r_{\rm max}=100a$ finding minimal changes in the surface density
distributions, accretion rates, and disc lopsidedness presented below.

From our set of $8$ (fiducial) $+16$ (parameter study) $=24$ CBD
simulations (Table \ref{Table:sims}), we use the following two
diagnostics to track the onset of the $q\sim 0.04$ cavity transition.
\begin{enumerate}
\item \textbf{Amplitude of accretion-rate variability:} To emphasize
  the change from steady-state to strongly-fluctuating solutions
  across the CBD phase transition, we compute the standard deviation
  from the mean accretion rate measured separately onto the individual
  BHs, as well as the total accretion rate onto both BHs,
\begin{eqnarray}
\delta \dot{M}^{n} &=& \sqrt{\frac{1}{N-1} \sum^N_j \left(\dot{M}^n_j -   \left<\dot{M^n}\right>_t \right)^{2}} \nonumber \\
\dot{M}^n(t) &=& \sum_{k} \frac{3}{2} \Sigma_k (t) (r^n_k)^{-2} \nu \qquad k \mid r^n_k \leq r^n_{\rm acc}.
\label{Eq:Mdot}
\end{eqnarray}
In the top equation $j$ denotes the
$j^{\rm th}$ timestep out of N total timesteps and $\left< \cdot \right>_{t}$ is the average over the entire time interval. In the bottom equation, $k$ denotes the $k^{\rm th}$ cell within the sink radius and the summation is over all cells within the sink radius. In both equations $n$ denotes the $n^{\rm th}$ binary component.
\item \textbf{Disc Lopsidedness:} To measure the lopsidedness of the
  cavity we measure the quantity
\begin{equation}
\epsilon = \left< \left< \frac{ \left| \left< \Sigma v_r \mbox{e}^{i \phi} \right>_{\phi} \right| }{\left< \Sigma v_{\phi}\right>_{\phi}}\right>_{r} \right>_{t},
\label{Eq:edsk}
\end{equation}
where $\left< \cdot \right>_{\phi}$ denotes an azimuthal average, $\left< \cdot \right>_{r}$ 
denotes a radial average from $r=a$ to the edge of the simulation domain, $\left< \cdot \right>_{t}$ 
denotes a time average over an integer number of binary orbits,
and $| \cdot |$ is the magnitude of a complex number. The only non-zero 
contributions to Eq. (\ref{Eq:edsk}) are from the components of $\Sigma v_r$ 
which are proportional to $A\cos{(\phi - \phi_A)} + B\sin{(\phi - \phi_B)}$, 
for arbitrary constants $A$, $B$, $\phi_A$, and $\phi_B$. Hence $\epsilon$ 
measures the lopsidedness of the disc. Note that $\epsilon$ is often referred to as the 
disc eccentricity \citep[{\em e.g.},][]{MacFadyen:2008,Farris:2014}.
\end{enumerate}
The above diagnostics are time-averaged over the final 100 orbits of
the simulation, for which a quasi-steady state has been achieved.
\\

\begin{table}
\begin{center}
\begin{tabular}{ c | c | c }
        $ q\equiv M_s/M_p   $                       & $\mathcal{M}$    & $\nu$ 		  \\
                   \hline 
0.01, 0.025, 0.075, 0.1                                      & 10      & $10^{-3}$      \\
 0.01, 0.025, 0.075, 0.1                                     & 10      & $10^{-4}$    \\
0.001, 0.01, 0.025, 0.03, 0.05, 0.075, 0.1, 1.0                 & 20      & $2.5 \times 10^{-5}$    \\
0.01, 0.025, 0.075, 0.1                                      &  30      & $10^{-3}$   \\
0.01, 0.025, 0.075, 0.1                                      &  30      & $10^{-4}$    
 \end{tabular}
\caption{Parameters for the simulations run in this study. The
  columns, from left to right, are the binary mass ratio, the orbital
  Mach number, and the coefficient of kinematic viscosity.}
\label{Table:sims}
\end{center}
\end{table}

\begin{figure*}
\begin{center}$
\begin{array}{c c c} \hspace{-0.5cm}
\includegraphics[scale=0.31]{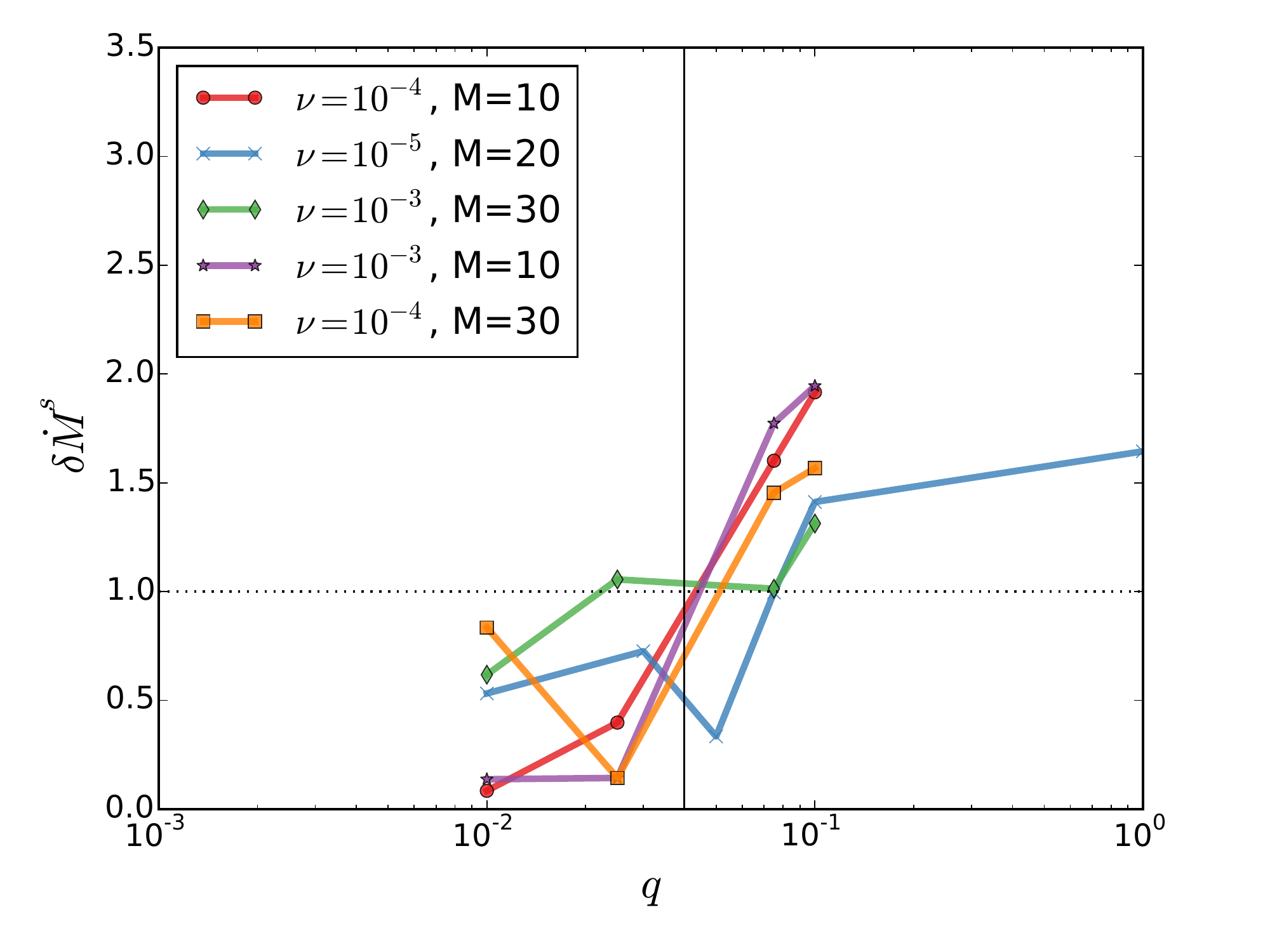} & \hspace{-0.5cm}
\includegraphics[scale=0.31]{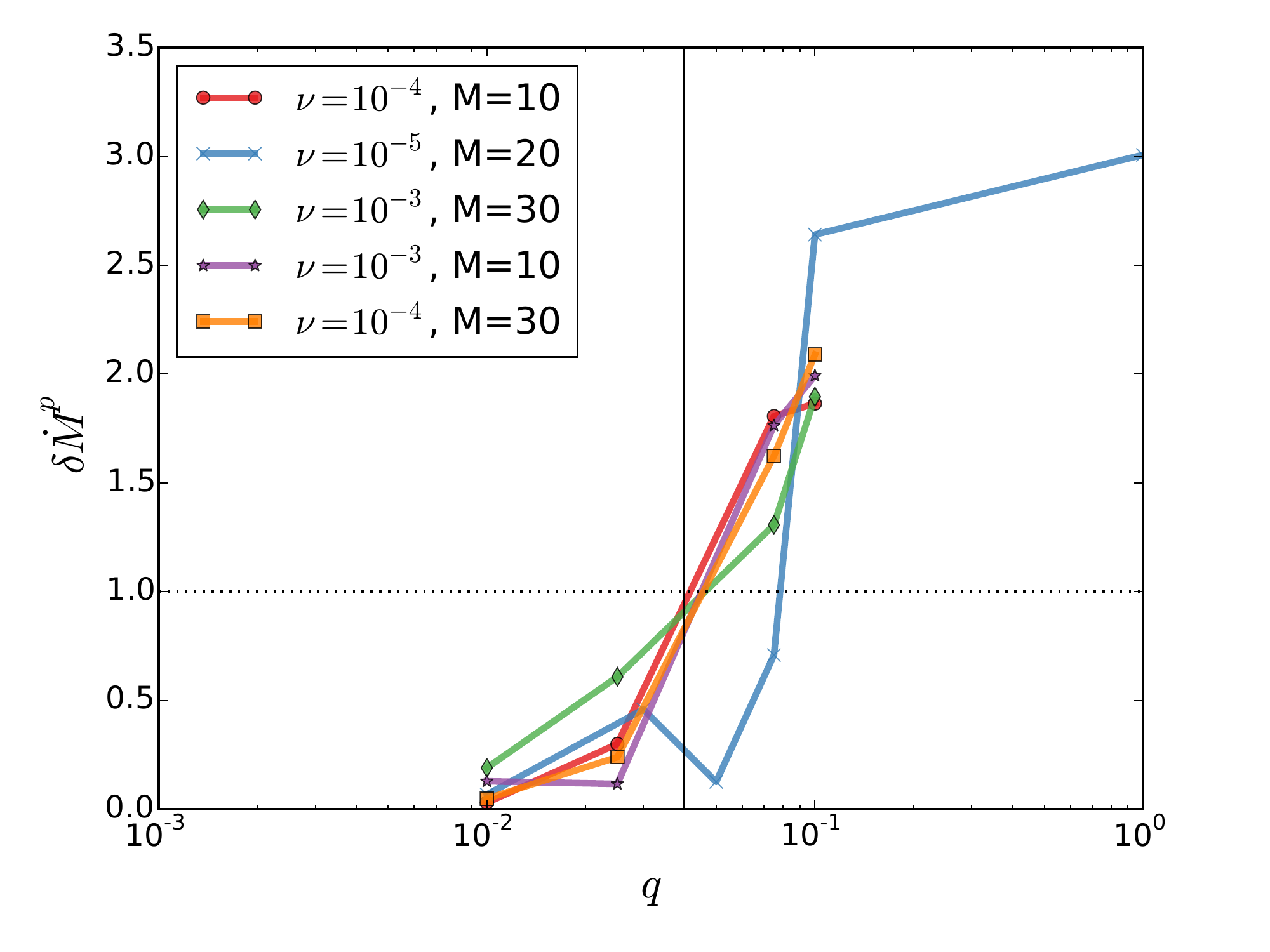} & \hspace{-0.5cm}
\includegraphics[scale=0.31]{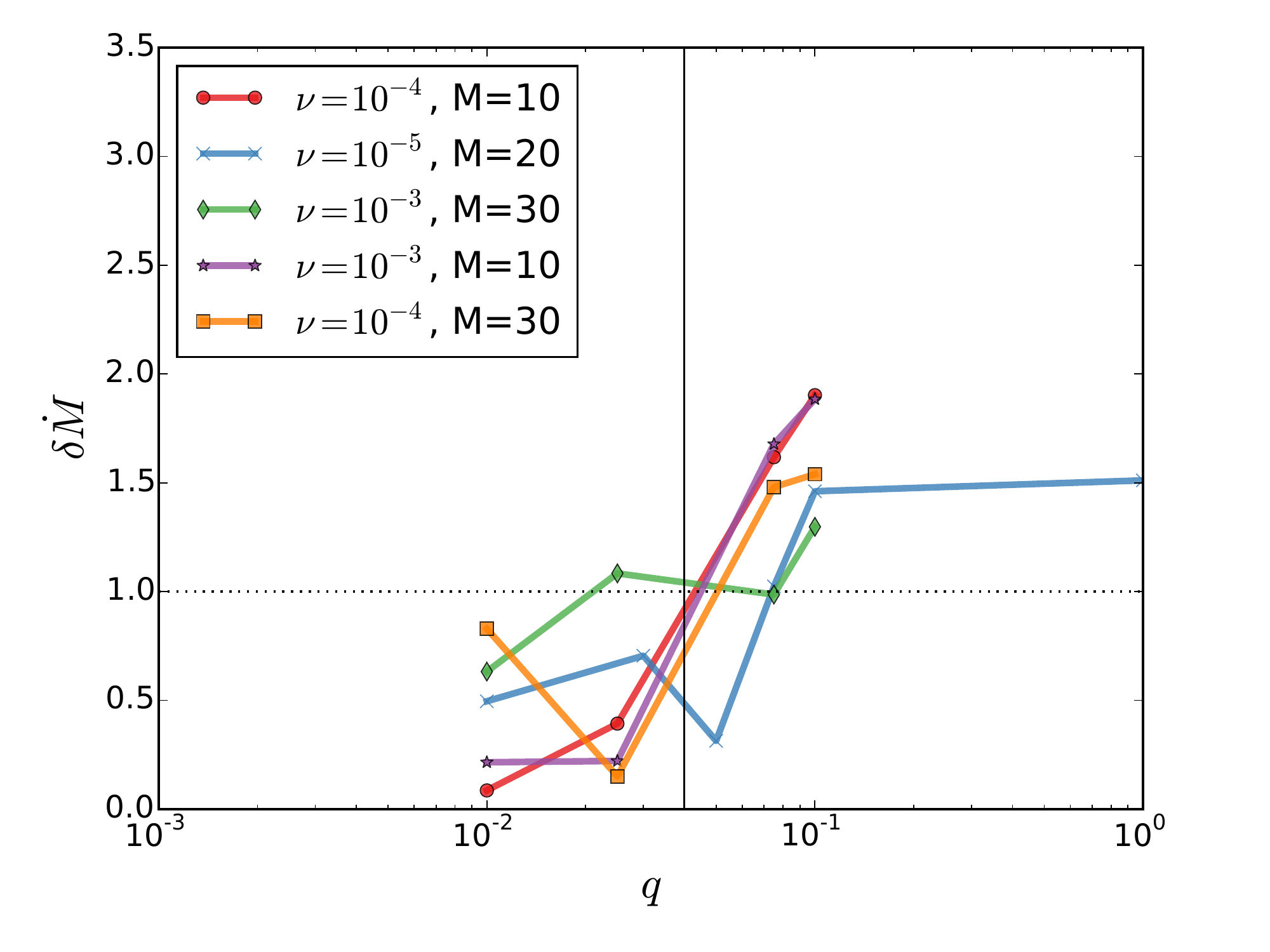}
\end{array}$
\end{center}
\caption{The standard deviation of $\dot{M}$ onto the secondary
  (left), primary (center), and both (right) BHs, computed over the
  final 100 orbits of each simulation. The vertical solid line is drawn at the 
  $q=0.04$ transition, the dashed horizontal line is the mean of the standard
  deviation of each data set.}
\label{Fig:Mdot}
\end{figure*}

Figure \ref{Fig:Mdot} plots the accretion rate variability onto the
secondary $\delta \dot{M}^s$ (left), primary $\delta \dot{M}^p$
(central), and both $\delta \dot{M}$ (right) BHs, as a function of
binary mass ratio for each set of disc parameters. Because there is a
scatter in the magnitude of accretion variability across the range of
different disc parameters (notably, the magnitude of variability
varies with viscosity), we normalize each set of mass ratios for a
given set of disc parameters by the average $\delta \dot{M}$ over all
$q$ (excluding the extreme values $q=0.001$ and $q=1$ for the fiducial
case). First notice that the left and right panels of Figure
\ref{Fig:Mdot}, $\delta \dot{M}^s$ and $\delta \dot{M}$, are very
similar. This is because over the range of mass ratios probed, the
accretion rate onto the secondary, and any resulting variation, is
larger than that onto the primary (in agreement with
\citealt{Farris:2014}). Both of these panels show a clear trend in
increasing accretion variability across the $q=0.04$ transition, for
all sets of disc parameters.

While observationally interesting, relative accretion rates measured
onto the secondary (and hence the total accretion rate onto both black
holes) may be a less robust diagnostic for the $q=0.04$ CBD transition
than accretion rates measured onto the primary. The transition
depicted by the central panel of Figure \ref{Fig:Mdot} - for the
standard deviation of accretion rates onto the primary - looks sharper
because the accretion rate variability settles down to small values
below the $q=0.04$ transition. The same measurements, for accretion
rate variability onto the secondary (and hence the total accretion
rate variability), do not settle to as tight a range of small
values. This effect may be tied to how accretion rates are measured in
the simulation; for small mass ratios, accretion is measured in the
small region within the Hill sphere of the secondary, and rates
measured in the simulations may be sensitive to this sink size and the
resolution. For example, for fiducial disc parameters, the $q=0.001$
and $q=0.025$ cases exhibit anomalously large variability onto the
secondary, which is omitted from Figure \ref{Fig:Mdot}.\footnote{For
  fiducial disc parameters the $q=0.025$ disc also exhibits
  anomalously large accretion variability onto the primary and should
  be investigated in future studies.} Future work should examine the
dependence of accretion rates on sink size, accretion prescription,
and resolution around the secondary. These considerations are,
however, less important for the primary.

The central panel of Figure \ref{Fig:Mdot}, displaying accretion
variability onto the primary, exhibits the most striking depiction of
the transition. This panel shows a sharp increase in variability
amplitude across the $q=0.04$ transition for all disc parameters. An
increased magnitude of accretion variability onto the primary is in
agreement with the earlier observation that the stream feeding the
primary BH becomes significant at the $q=0.04$ transition. Because the
second stream could be necessary for generating the lopsided cavity,
$\delta \dot{M}^p$ acts as an excellent diagnostic for the $q=0.04$
transition.

\begin{figure}
\begin{center}
\includegraphics[scale=0.4]{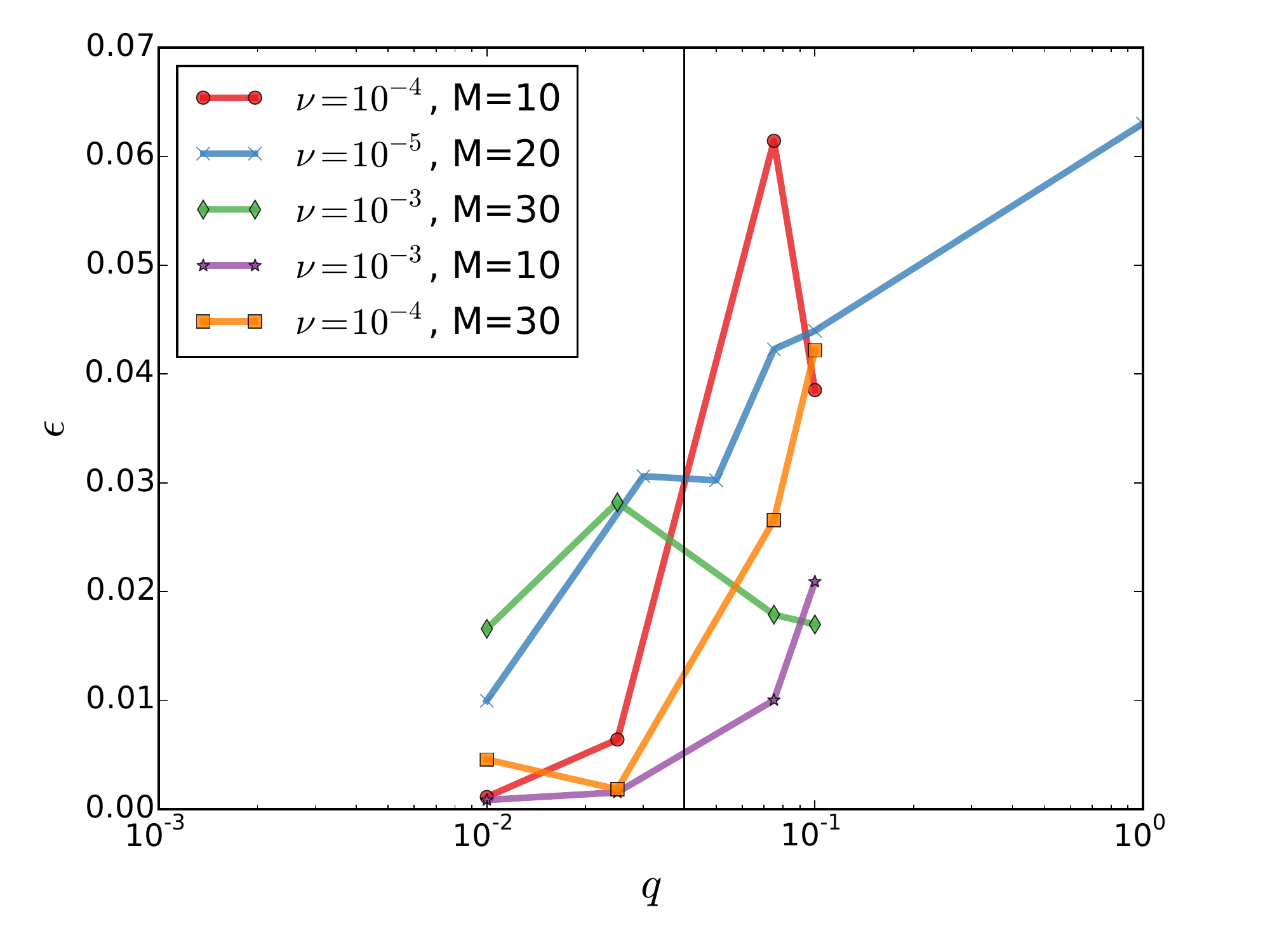} \\
\end{center}
\caption{The lopsidedness, Eq. (\ref{Eq:edsk}), of the circumbinary
  disc, spatially averaged over the region outside of the binary
  orbit, $r>a$, and time averaged over the final 100 orbits of each
  simulation. The vertical sold line denotes $q=0.04$.}
\label{Fig:edsk}
\end{figure}

\cite{MacFadyen:2008, ShiKrolik:2012:ApJ, Noble+2012, DHM:2013:MNRAS,
  Farris:2014,Farris:2015:Cool,Farris:2015:GW, ShiKrolik:2015} have
shown that an equal-mass binary can create a lopsided CBD (referred to
as eccentric in \cite{MacFadyen:2008} and
\cite{Farris:2014}). \cite{Farris:2014} measured the disc lopsidedness
for a range of binary mass ratios, finding a sharp transition to
lopsided discs for $q\gsim0.05$ (see Figures 5 and 6 of
\citealt{Farris:2014}). Figure \ref{Fig:edsk} plots the time averaged
disc lopsidedness, Eq. (\ref{Eq:edsk}), measured at radii $r>a$, as a
function of binary mass ratio. We find a sharp increase in disc
lopsidedness near $q=0.04$ for all disc viscosities and pressures
except for the high viscosity $\nu_0=10^{-3}$, low pressure case
$\Mach=30$.

\begin{figure}
\begin{center}$
\begin{array}{c c}
\hspace{-20 pt}
\includegraphics[scale=0.25]{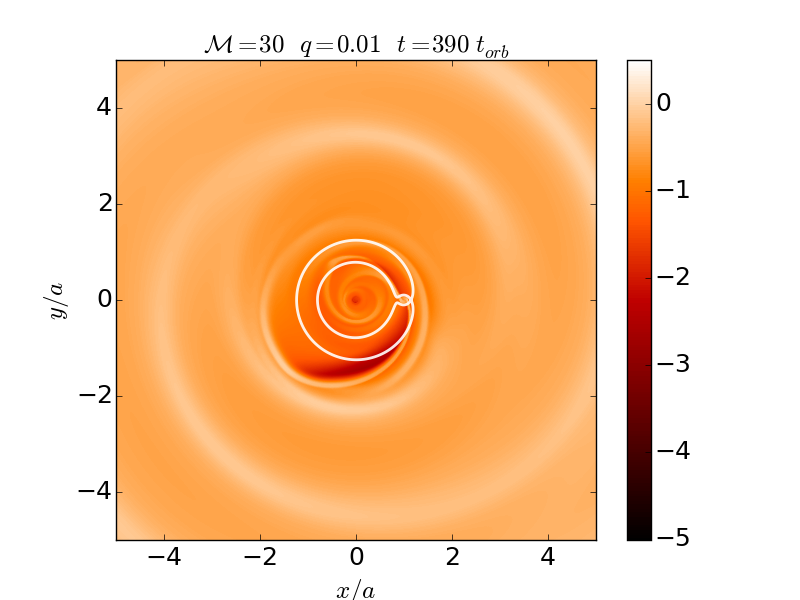}& \hspace{-30 pt}
 \includegraphics[scale=0.25]{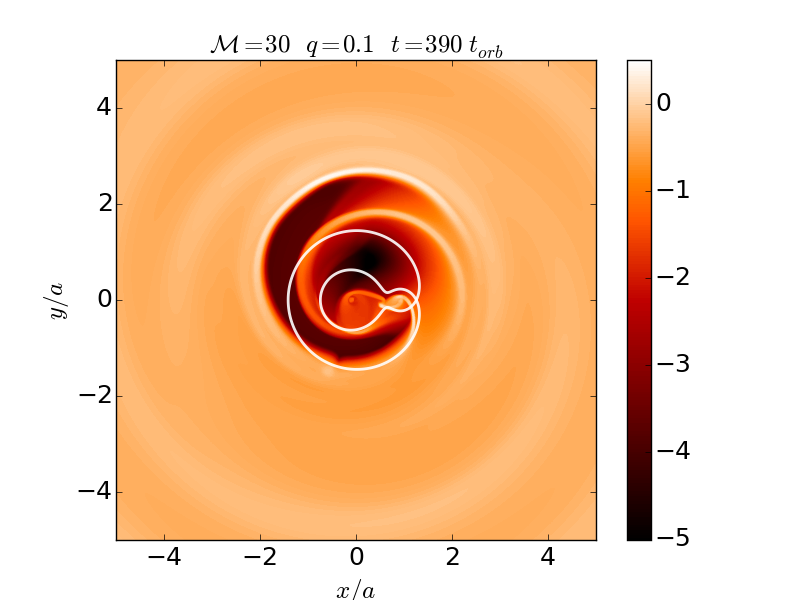} 
\end{array}$
\end{center}
\caption{Snapshots of the surface density distribution (shown in units
  of the unperturbed value, with a logarithmic colour scheme) from
  hydrodynamical simulations for a disc with $\Mach = 30$, and
  constant coefficient of kinematic viscosity $\nu =10^{-3} a^2_0
  \Omega_{\bin}$. For these high-viscosity, low-pressure simulations,
  we find an asymmetric-disc shape even for small mass ratio binaries (left). The
  transition to a time-dependent, lopsided cavity at $q\sim0.04$ still
  takes over for larger mass ratio binaries (right).}
\label{Fig:n3M30}
\end{figure}

We display the surface density of the $q=0.01$, $\nu_0=10^{-3}$,
$\Mach=30$ disc in the left panel of Figure \ref{Fig:n3M30}. In this
case, strong shocks form intermittently as the spiral arm connecting
to the secondary propagates outwards into the disc. These shocks lead
to the formation of a slightly lopsided disc even for the $q=0.01$
case. We note that this asymmetry could be caused by the difference in
instability times derived for the viscous R3Bp in the Appendix (see
Figure \ref{Fig:Lam_L4L5}). However, we leave an investigation for
future work

In addition to the possible viscous instability explained in the
Appendix, there are other mechanisms which may make the CBD lopsided
and induce variable accretion. \cite{KleyDirksen:2006} find that a
spatial gap asymmetry can be induced even for mass ratios as low as
$q=0.003$. This is only found to occur when viscosity is low enough
for a deep gap to be cleared. The explanation of
\cite{KleyDirksen:2006} uses the work of \cite{Lubow:1991:Model,
  Lubow:1991:Sim}; a deep gap mitigates the eccentric co-rotation
resonances in the disc which act to damp disc lopsidedness while
eccentric Linblad resonances (eLRs) act to grow disc
lopsidedness. This asymmetry growth due to eLRs is not likely the
mechanism causing the $q=0.04$ phase transition. Figures
\ref{Fig:Mdot} and \ref{Fig:edsk} show that the transition to a
lopsided cavity occurs around $q=0.04$ even for large viscosities and
that there is no trend of a decreasing critical mass ratio with
decreasing viscosity as the \cite{KleyDirksen:2006} mechanism would
predict. Additionally, \cite{ShiKrolik:2012:ApJ} show that the growth
timescale of asymmetry for the $q=1$ binary does not match the growth
timescale expected from the argument of
\cite{KleyDirksen:2006}. Future work should measure the growth rate of
lopsidedness for multiple mass ratios in order to dis-entangle the
mechanism by which the low mass ratio systems of
\cite{KleyDirksen:2006} become asymmetric (presumably eLRs) and the
mechanism which causes the phase transition studied here (presumably
orbital instability).

Regardless of hydrodynamic processes which may act to grow a lopsided,
time-variable disc below the $q=0.04$ transition, the present study
shows that a robust transition in CBD structure dominates for $q\gsim
0.04$ over a wide range of hydrodynamic disc parameters.

\section{Discussion and Summary}
Circumbinary accretion discs exhibit a phase transition above a binary
mass ratio of $q \simeq 0.04$. The transition signifies a loss of
spatial and time-translation symmetry in the CBD. The density
structure changes from an annular low-density gap near the orbit of
the secondary ($q\lsim 0.04$), to a lopsided cavity ($q\gsim
0.04$). It also marks a transition from steady-state to
strongly-fluctuating behaviour. We conjecture that this transition is
closely tied to the loss of stable orbits in the binary co-orbital
region of the restricted three-body problem (R3Bp) for $q>0.04$.
When stable co-orbital orbits connect the outer, circumbinary disc 
to the inner, circumprimary disc, steady-state gapped solutions are 
realized. When no such stable co-orbital orbits exist, accretion 
streams violently impact the inner mini-discs and the outer 
circumbinary disc, leading to fluctuating lopsided cavity solutions. We 
employ the R3Bp as well as 2D viscous hydrodynamical simulations to 
investigate the CBD transition.

We show that the change in density morphology, from annular gaps to
cavities, in circumbinary accretion discs can be largely captured by
the spatial restriction of test particles imposed by conservation of
the Jacobi constant in the R3Bp. To quantify the limitations of the
R3Bp and to more closely compare to hydrodynamical simulations, we
extend the R3Bp analysis to include the effects of pressure and
viscosity. To estimate the effects of pressure in the disc, we compare
the Jacobi constant with the closely related Bernoulli constant and
derive a maximum disc aspect ratio (minimum Mach number) necessary for
pressure forces to overcome the gravitational barrier of the
binary. For binaries with $q>0.04$ this only occurs for discs which
are no longer thin.

To study the effects of viscosity on the R3Bp and investigate the
relation of the CBD phase transition to orbit stability, we add a
viscous force to the R3Bp equations and perform a linear-stability
analysis on orbits around the L4 and L5 equilibrium points in the
presence of this force. We find that the viscous R3Bp shares a similar
orbital stability transition which occurs very near the classical
stability mass ratio of $q=0.04$ over multiple magnitudes of viscosity
(see the Appendix), consistent with our hydrodynamical parameter
study.  We also find that viscous forces break a symmetry between the
stability of the leading and trailing triangular Lagrange points
(L4/L5) of the classical R3Bp. This may be related to the growth of
asymmetry in CBDs, though we save such a study for future work.

The effects of both viscosity and pressure on the CBD transition are
studied via 2D viscous hydrodynamical simulations. These show that
changing the disc viscosity by a factor of 40 and the disc pressure by
a factor of 9 leaves the critical mass ratio largely unaffected.  The
hydrodynamical simulations also provides further evidence for a
mechanism by which orbital instability seeds the transition. For
$q\lsim 0.04$ disc particles in the binary co-orbital region stably
oscillate on horseshoe-like orbits, while for $q \gsim 0.04$,
particles are flung out of the co-orbital region into the outer disc
(Figure \ref{Fig:PS}).

We note that, in addition to loss of L4/L5 orbital stability, there is
another dynamical transition which occurs near $q=0.04$ in the
R3Bp. With the goal of determining the truncation of CBDs,
\cite{RP:Excretion:1981} studied the intersection and radial stability
of test particle orbits around a circular binary with arbitrary mass
ratio. They posit that the innermost stable orbits of a CBD are set
either by orbit intersection, or instability to radial perturbations
of a Keplerian orbit around the binary. They find that, for binaries
with $q \lsim 0.01$, orbit intersection truncates the CBD, for $0.01
\leq q \leq 0.05$ the inner edge of the disc becomes marginally
unstable before orbit intersection becomes important, and for
$q>0.05$, violent instability is responsible for disc truncation. In
this paper, we have shown evidence that links the L4/L5 orbital
instability to the observed CBD transition. The additional radial
instability found by \cite{RP:Excretion:1981} in the particle limit,
near the inner edge of the CBD, also occurs near q=0.05. However, our
hydrodynamical simulations do not show strong fluctuations in the
inner edge of the CBD for $q \gsim 0.05$. Rather, the variability in
the accretion rates we observe appears to track the unstable behavior
of the gas near L4/L5 (see Figure \ref{Fig:PS} and corresponding
movies).

In accordance with the simplicity of the circular R3Bp and our
numerical simulations, we have restricted our analysis to systems
consisting of isothermal and adiabatic thin discs surrounding a binary
on a fixed circular orbit. A binary on an eccentric orbit exhibits
loss of co-orbital stability at mass ratios smaller than $q=0.04$ of
the circular case \citep{eR3BStability:2013}. Supersonic gas dynamics
in the vicinity of the binary will likely invalidate the assumption of
an isothermal gas disc. A massive disc will move the binary; for mass
ratios above the transition, the binary may become eccentric
\citep{Cuadra:2009, Roedig:2011:eccevo,Roedig:2012:Trqs} and migrate
differently than the often assumed Type II rate
\cite[\textit{e.g.}][]{HKM09,DottiMM:2015}. Future work should address
the change of binary orbital parameters due to the change in CBD
structure. Other important physics not included in this work may also
impact the $q=0.04$ transition. Including the vertical disc dimension,
magnetic fields and the magneto-rotational instability, and radiative
feedback from accretion could all affect the onset and behaviour of
the transition described here.

Future work on CBD structure may find insight into more general
binary+disc systems from extensions of the circular R3Bp for a binary
with time-dependent separation \citep{SchnitL4L5:2010}, a mis-aligned
disc \citep{ErwinSparke:1999}, or non-zero eccentricity
\citep{PSA:2005}.

Within the limitations of this study, we have identified a dramatic
transition occurring in circumbinary discs and offered an intriguingly
simple origin; further work should clarify whether it survives the
additional physical effects mentioned above.

The $q=0.04$ CBD transition is relevant to MBHB+disc
systems. Accreting MBHBs below the critical mass ratio will prove
difficult to detect in time-domain electromagnetic surveys due to
accretion variability alone. For such low mass, steadily-accreting
MBHBs, other mechanism could cause variability, such as a precessing
jet, disc instabilities, or relativistic Doppler boost in the case of
compact binaries \citep{PG1302Nature:2015b}.  Because of the drastic
change in disc lopsidedness across the $q=0.04$ transition, there may
be an equally drastic change in the binary orbital parameters. This
would result in a connection of the binary mass ratio with binary
eccentricity and migration rate. This could affect gravitational wave
detection rates as well as waveforms.

The $q=0.04$ CBD transition is also relevant for proto-planetary
systems. The formation of planets around a binary system may progress
differently for a proto-planetary disc around a brown dwarf and main
sequence pair \citep[\textit{e.g},][]{q0p02MSBD:2014,
  StellMassRatDist:2015, 7StellarEMRs:2015} then for a near equal mass
binary. Though theoretically disfavoured \citep{PayneLodato:2007},
there would be important consequences for the formation of planetary
systems around brown dwarfs which contain a large ($\gsim M_J$)
planet. It will be interesting to look for differences in planetary
populations around binaries above and below $q\sim 0.04$ if they are
discovered in upcoming searches \citep[\textit{e.g.}][]{Triaud:2013,
  TESS:2014}.

\section*{Acknowledgments}
The authors thank Jeffery J. Andrews and Adrian Price-Whelan for
useful discussion of the two- and three-body problems, and Ganesh
Ravichandran for writing an initial version of a code used for
analysis of the hydrodynamical parameter study. The authors also thank
Roman Rafikov, Jeno Sokolowski, and Scott Tremaine for useful
discussions. We acknowledge support from a National Science Foundation
Graduate Research Fellowship under Grant No. DGE1144155 (DJD) and NASA
grant NNX11AE05G (ZH and AM).

\appendix
\section{Viscous, restricted three-body problem}
\label{AppendixA}
We follow \cite{Murray:1994} and \cite{MD:SSD} to add an external
viscous force $\mathbf{F}(x,y,\dot{x},\dot{y})$ to the R3Bp equations
of motion. In the rotating frame,
\begin{equation}
\begin{array}{c}
\ddot{x} - 2\Omega\dot{y} = \frac{\partial{U}}{\partial{x}} + F_{x}  \nonumber \\ \nonumber \\ 
\ddot{y} + 2\Omega\dot{x} = \frac{\partial{U}}{\partial{y}} + F_{y},
\label{Eqmotion}
\end{array}
\end{equation}
where $\Omega$ is the frequency of the rotating frame and $U(x,y)$ is
the Roche potential with sign convention taken to be consistent with
\cite{MD:SSD}.

For the external force we use the force associated with the $r-\phi$
component of the viscous stress tensor in a Keplerian flow. In the
inertial frame this force (per unit mass) is
\begin{equation}
\begin{array}{l}
F_{\hat{\phi}} = \frac{\partial_{r} \sigma^{\hat{r} \hat{\phi}}}{\rho} = -\partial_r\left[ \nu r \partial_r \Omega_{\rm disc} \right] = -\frac{9}{4}  \nu \frac{v_{\phi}}{r^2}, \nonumber
\end{array}
\end{equation}
where we assume that the angular velocity of the flow is Keplerian
around the binary mass $M$, and that the kinematic coefficient of viscosity $\nu$ and
density $\rho$ are constants. In the last line we relabel $v_{\phi}
\equiv \sqrt{GM/r}$ . By transforming to Cartesian coordinates and
then moving into the rotating frame we find
\begin{equation}
\begin{array}{l}
F_{x} = -F_{\hat{\phi}} \frac{y}{r} \nonumber \\ \nonumber \\ 
F_{y} = F_{\hat{\phi}} \frac{x}{r}.
\end{array}
\end{equation}
Writing these in terms of the rotating frame coordinates and
velocities,
\begin{equation}
\label{Eq:VscF}
\begin{array}{l}
F_{x} = -\frac{9}{4}\frac{\nu}{x^2+y^2}(\dot{x} - y \Omega) \nonumber \\ \nonumber \\ 
F_{y} = -\frac{9}{4}\frac{\nu}{x^2+y^2}(\dot{y} + x \Omega),
\end{array}
\end{equation}
which follows from our choice of azimuthal velocity profile
$v_{\phi}(r)$ in the inertial frame of the binary so that $\dot{x} =
-(v_{\phi} - r\Omega)y/r$ and $\dot{y} = (v_{\phi} - r\Omega)x/r$.

As with pressure forces (\S \ref{Pressure}), viscous forces also destroy conservation of
the Jacobi constant. $C_J$ changes at a rate
\begin{equation}
\dot{C_J} = \frac{9}{2} \frac{\nu}{x^2+y^2} \left[ \dot{x}^2  + \dot{y}^2 + \Omega (x\dot{y} - y \dot{x}) \right].
\end{equation}
Comparing the change in $C_J$ due to viscosity to the change due to
pressure over $N_{\rm orb}$ orbital times gives,
\begin{equation}
|\Delta C_J |_{\nu} = \alpha N_{\rm orb} \frac{5}{9 \pi} |\Delta C_J |_{P},
\end{equation}
where we have used Eq. (\ref{Eq:CB}) and the first of
Eqs. (\ref{dP_iso}) for the RHS and assumed a constant coefficient of
kinematic viscosity $\nu = \alpha/\Mach^2 a^2 \Omega$ with $\alpha <
1$. In the case that $N_{\rm orb}$ is a gap opening time, which we
approximate as the viscous time across a gap of size equal to the disc
scale height, we find
\begin{equation}
|\Delta C_J |_{\nu} = \frac{5}{27 \pi^2} |\Delta C_J |_{P} \sim  0.02 |\Delta C_J |_{P}.
\end{equation}
Hence for short timescales relevant for local density distribution
such as gaps, the pressure forces dominate in the decay of the Jacobi
constant. For longer timescales $N_{\rm orb} \gg 1/\alpha$ it is viscosity which dominates
the deviation from the purely gravitational problem.

To investigate the consequences of a viscous force for orbital
stability, we perform a linear-stability analysis for particles
perturbed from the analogue of the L4/L5 points in the viscous R3Bp
(\ref{Eqmotion}), with external force (\ref{Eq:VscF}). The location of
the new equilibrium points in the presence of the viscous force are
found by setting $\ddot{x} = \ddot{y} = \dot{x} = \dot{y} =0$ in the
equations of motion (\ref{Eqmotion}) and solving for the coordinates
$(x^{\dagger},y^{\dagger})$ of the new equilibrium
points. \cite{Murray:1994} finds solutions by Taylor expanding the
equations of motion around the $\mathbf{F}=0$ classical equilibrium
points $(x_0,y_0)$ and solving for the deviation $(\bar{x},\bar{y})$
from $(x_0,y_0)$. To linear order in $\bar{x},\bar{y}$ the new
equilibrium points are,
\begin{equation}
\begin{array}{l}
x^\dagger = x_0 - \frac{(1+q)}{q} \frac{F^*_x}{3 } \pm  \frac{(1+q)}{q} \frac{F^*_y}{3\sqrt3 } \nonumber \\ \nonumber \\ 
y^\dagger = y_0 \pm \frac{(1+q)}{q} \frac{F^*_x}{3\sqrt3 } -  \frac{(1+q)}{q} \frac{F^*_y}{9 }, 
\label{xyDag}
\end{array}
\end{equation}
where $+$ and $-$ denote L4 and L5, and $*$ denotes evaluation at the
classical Lagrange point $(x=x_0, y=y_0, \dot{x}=0, \dot{y}=0)$.

\begin{figure}
\begin{center}$
\begin{array}{c}
 \hspace{-30 pt}
\includegraphics[scale=0.45]{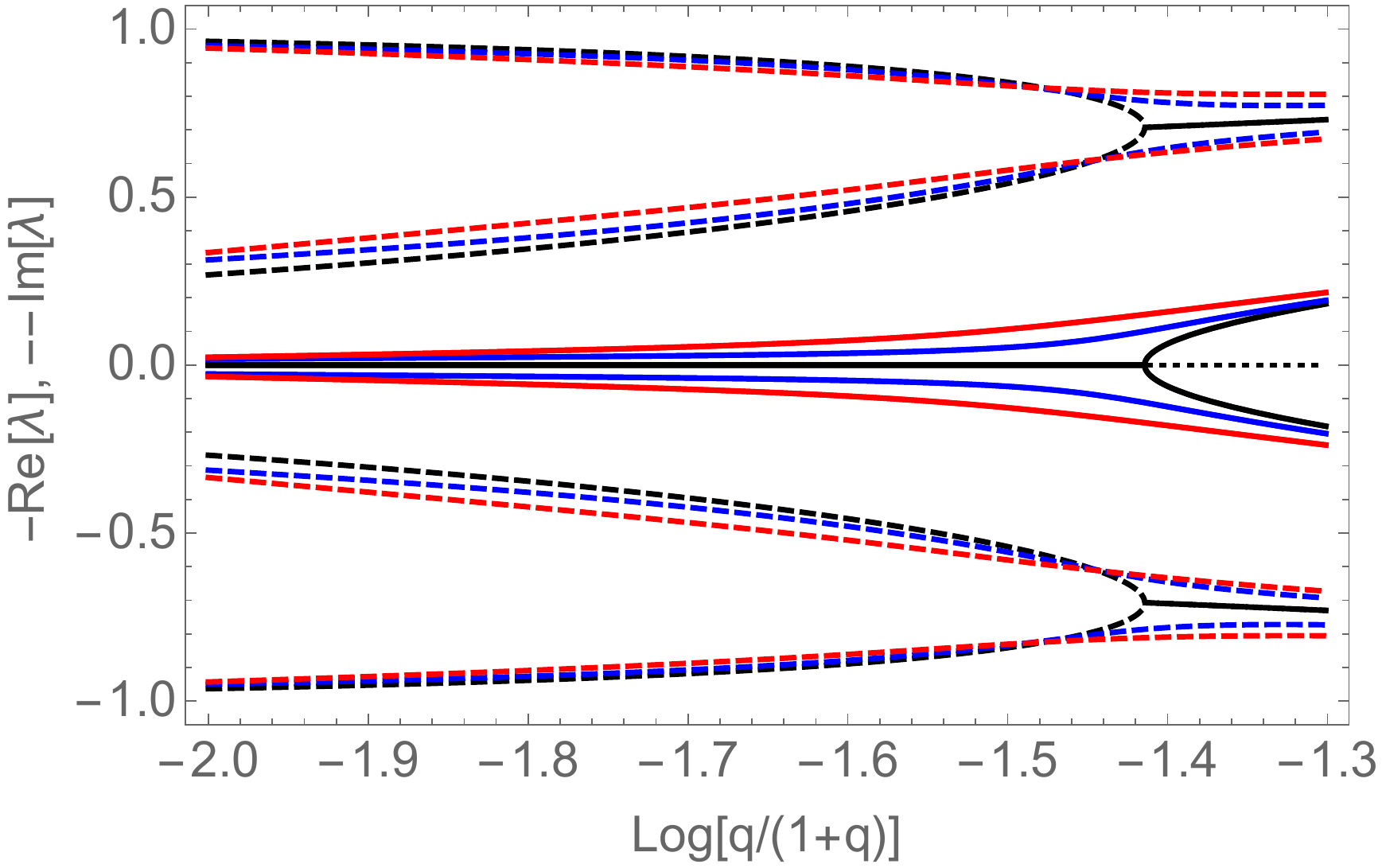} \\
\includegraphics[scale=0.535]{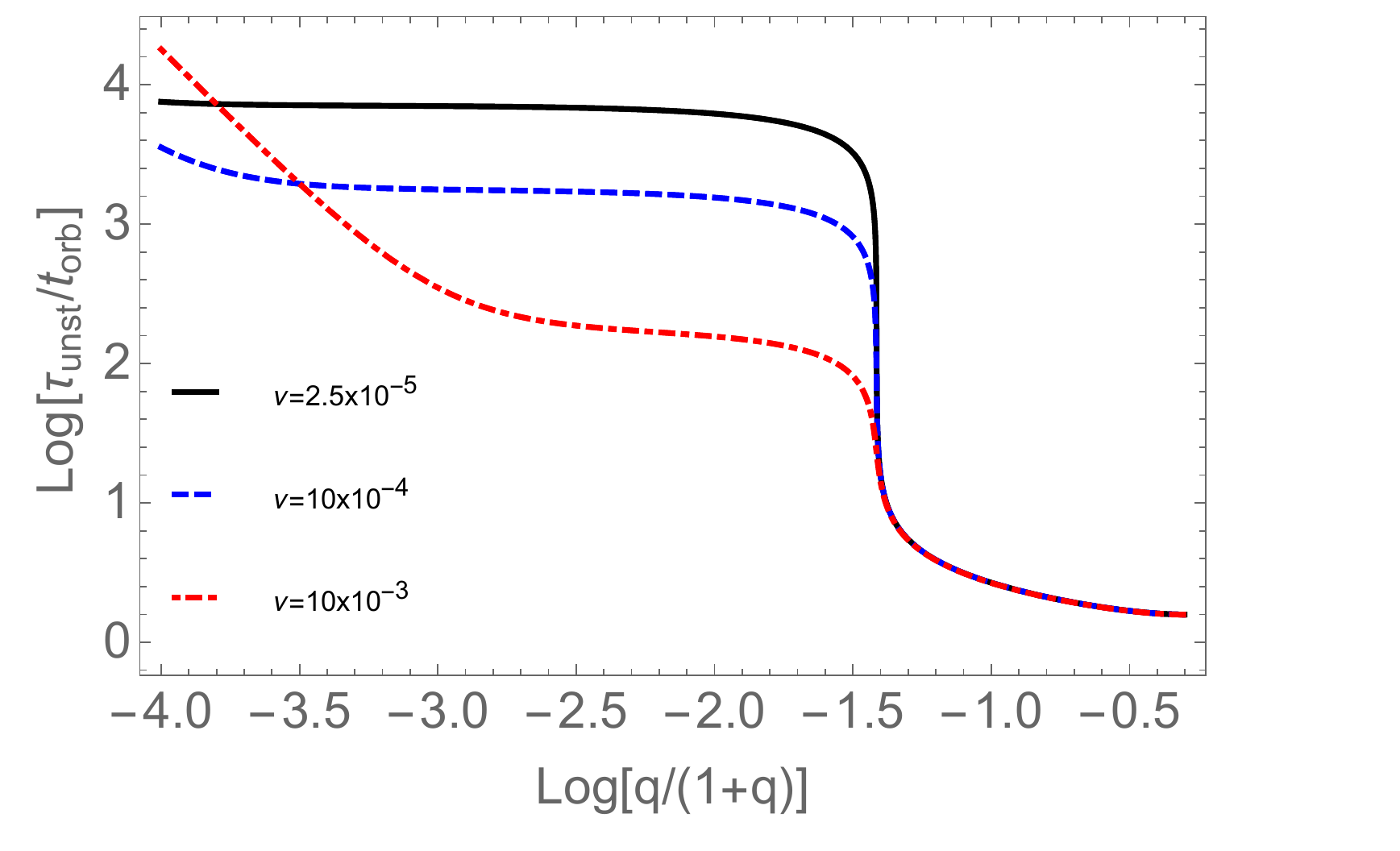}
\end{array}$
\end{center}
\caption{Eigenvalues of a linear-stability analysis of the L4 and L5
  points with an added viscous force. The top panel shows the real
  (solid lines) and imaginary (dashed lines) parts for different values of
  the constant coefficient of kinematic viscosity, $\nu=0.0$ (black),
  $\nu=0.005$ (blue), $\nu=0.01$ (red). The bottom panel displays
  instability timescales normalized to the binary orbital time.}
\label{Fig:Lam}
\end{figure}

\begin{figure}
\begin{center}$
\begin{array}{c}
 \hspace{-10 pt}
\includegraphics[scale=0.24]{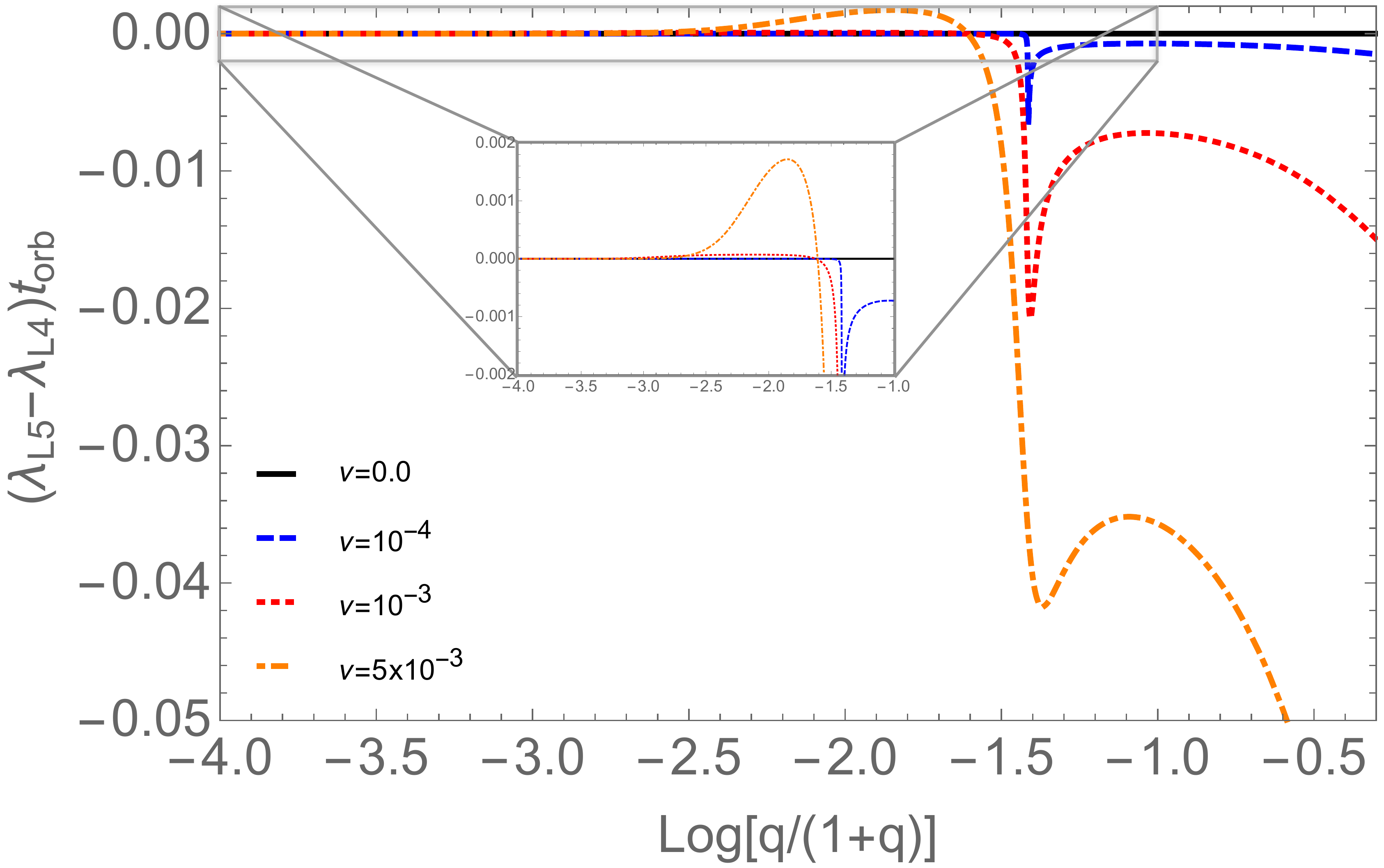} 
\end{array}$
\end{center}
\caption{The difference in orbital out-spiral of a test-particle
  perturbed from the L4 and L5 points. $\lambda_{\rm L4}$ and
  $\lambda_{\rm L5}$ refer to the maximum real part of the complex
  eigenvalues found from Eq. \ref{Eq:Lam}. The inset zooms in on the
  region below the $q=0.04$ transition where weak asymmetry between
  the L4 and L5 points exists. A value of $-0.01$ in this plot means
  that, after 100 binary orbital times, the difference in the final position of
  a particle perturbed from L4 is an e-fold farther from its starting
  position than a particle identically perturbed from L5. It is likely 
  that a higher order perturbation in the mass ratio is required to 
  capture the return of $\lambda_{\rm L5}-\lambda_{\rm L4}=0$ for $q=1$.}
\label{Fig:Lam_L4L5}
\end{figure}

Next assume a solution for the motion of a test particle perturbed
from L4/L5 to be of the form,
\begin{equation}
\label{ansatz}
x(t) = x^{\dagger} + X_0 \rm{e}^{\lambda t} \qquad y(t) = y^{\dagger} + Y_0 \rm{e}^{\lambda t}.
\end{equation}
Substituting this ansatz into the equations of motion (\ref{Eqmotion})
and keeping terms to linear order in the displacement,
$X_0\rm{e}^{\lambda t}$ and $Y_0\rm{e}^{\lambda t}$,
\cite{Murray:1994} finds a set of simultaneous linear equations for
the displacement from equilibrium with characteristic equation for the
eigenvalues $\lambda$,
\begin{equation}
\lambda^4 + a_3 \lambda^3  + (1+a_2)\lambda^2 + a_1 \lambda + \frac{27}{4}\frac{q}{(1+q)^2} + a_0 = 0,
\label{Eq:Lam}
\end{equation}
which assumes that the viscous force is small by neglecting terms of
$\mathcal{O}(\nu^2)$.  The $a_i$ coefficients can be written in terms
of derivatives of the external force evaluated at the new equilibrium
points. In the case of our viscous force
\begin{equation}
\begin{array}{l}
a_0 =  \frac{27}{4}\nu\frac{ \pm \sqrt{3} (x^{\dagger})^2 + 2 x^{\dagger} y^{\dagger} \mp \sqrt{3} (y^{\dagger})^2 }{2  \left[(x^{\dagger})^2  +   (y^{\dagger})^2\right]^2}   \nonumber \\
a_1 =  \frac{27}{4}\frac{\nu }{ (x^{\dagger})^2  +    (y^{\dagger})^2}  \nonumber \\
a_2 = 0   \nonumber \\
a_3 =  - \frac{9}{2}\frac{\nu }{ (x^{\dagger})^2  +    (y^{\dagger})^2},
\end{array}
\end{equation}
where $\pm$ and $\mp$ refer to the L4 and L5 points respectively,
$x^{\dagger}, y^{\dagger}$ are given by (\ref{xyDag}), and we have
dropped terms of order $q^2/(1+q)^2$, except in the constant term
(second to last term on the LHS of Eq. (\ref{Eq:Lam})) which gives the
stability criterion of $q<0.04$ in the classical R3Bp. Solving the
depressed quartic for $\lambda$ we find four complex solutions
which we plot in Figure \ref{Fig:Lam} for multiple values of the
viscosity. We make three observations, 
\begin{itemize}
\item For finite viscosity and $q>0$, there are no linearly stable
  orbits, only quasi-stable orbits, which are formally unstable but
  have long instability times (small Re$[\lambda]$). This can be
  seen from the solid lines in the top panel of Figure
  \ref{Fig:Lam}. The non-viscous case (black lines) has Re$[\lambda]
  =0$ for $q<0.04$, the classical result. For the higher-viscosity
  cases (blue and red lines), $|\rm{Re}[\lambda]|$ is large and
  Eqs. (\ref{ansatz}) show that orbits will decay or blow up.
\item If we define the critical mass ratio for linear quasi-stability
  to be the mass ratio where Re$[\lambda]$ has the largest derivative
  (hence changing quickly with $q$ from a long to short instability
  timescale), then the top panel of Figure $\ref{Fig:Lam}$ shows that
  this critical mass ratio becomes smaller for larger viscous
  forces. The instability transition is explored further in the bottom
  panel of Figure \ref{Fig:Lam} where we plot the instability
  timescale in units of the orbital time. The critical mass ratio is
  not very sensitive to the viscosity. The instability timescale is of
  order the viscous time below $q=0.04$ and of order an orbital time
  above $q=0.04$.
\item The symmetry between the L4 and L5 points is broken. Figure
  \ref{Fig:Lam_L4L5} plots the difference in maximum Re$[\lambda]$ between
  the L4 and the L5 points in units of the inverse binary orbital time. As expected, there
  is no difference in growth timescale for L4 vs. L5 for the
  non-viscous case (black line). The viscous (red, blue, and orange
  lines) cases, however, show a large difference in instability growth
  rate above $q=0.04$ and a weaker difference over a viscosity
  dependent mass ratio range below the $q=0.04$ transition. For
  smaller mass ratios, the L5 point is more unstable, and for larger
  mass ratios, the L4 point is more unstable. 
\end{itemize}

For completeness we include the oscillatory part of the linear
solutions (Im$[\lambda]$) in the top panel of Figure \ref{Fig:Lam}
(dashed lines). The different oscillation timescales are discussed for
the non-viscous R3Bp in \citep{Murray:1994}.

\begin{figure}
\begin{center}$
\begin{array}{c c }
 \includegraphics[scale=0.38]{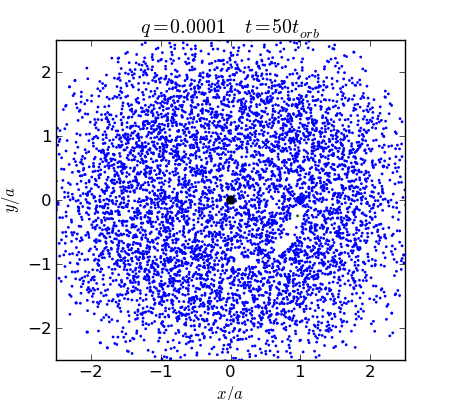} & \hspace{-15 pt}
\includegraphics[scale=0.38]{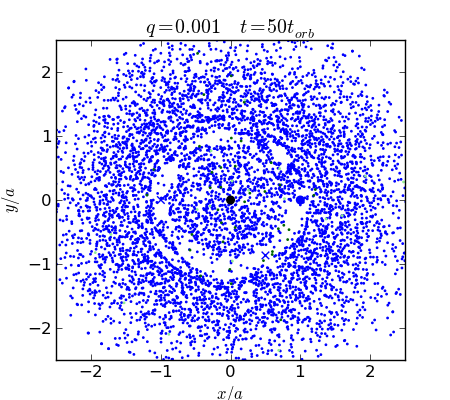} \\
\includegraphics[scale=0.38]{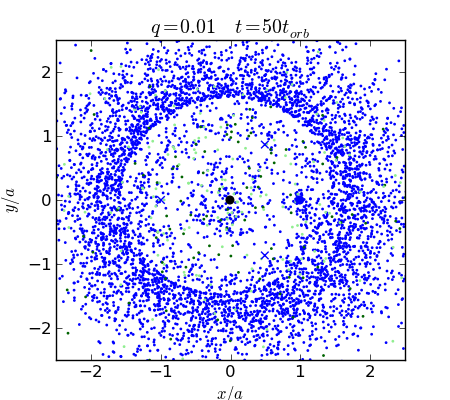} & \hspace{-15 pt}
\includegraphics[scale=0.38]{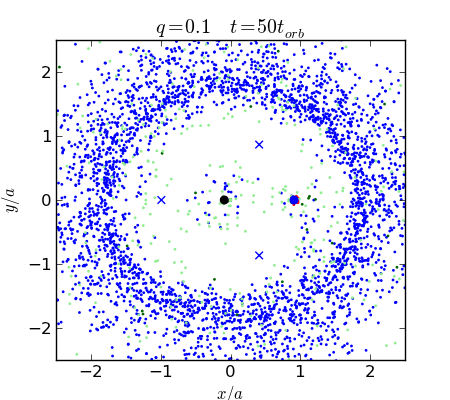} \\
 \includegraphics[scale=0.38]{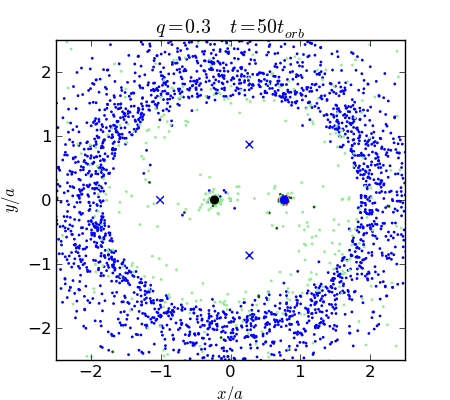} & \hspace{-15 pt}
 \includegraphics[scale=0.38]{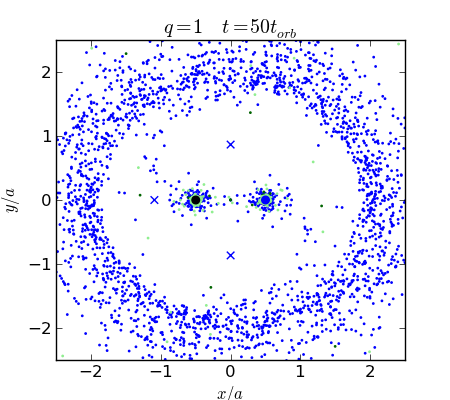}
\end{array}$
\end{center}
\caption{Same as Figure \ref{Fig:Int_Orb} except for $10^4$ 
particles obeying the viscous R3Bp 
Eqs. (\ref{Eqmotion}) and (\ref{Eq:VscF}) with constant $\nu =
0.001 a^2_0 \Omega_0$.}
\label{Fig:Int_Orb_Vsc}
\end{figure}

We also integrate the full viscous R3Bp in Figure
\ref{Fig:Int_Orb_Vsc}. Streams feeding the binary components exist
even after $50$ orbits in the case where viscous forces are allowed
to destroy conservation of the Jacobi constant.

\bibliographystyle{mnras}
\bibliography{Dan}

\end{document}